\documentclass[fleqn,usenatbib]{mnras}
\usepackage{newtxtext,newtxmath}
\usepackage[T1]{fontenc}
\DeclareRobustCommand{\VAN}[3]{#2}
\let\VANthebibliography\thebibliography
\def\thebibliography{\DeclareRobustCommand{\VAN}[3]{##3}\VANthebibliography}

\usepackage{amsmath}
\usepackage{graphicx}
\usepackage[dvipsnames]{xcolor}   
\usepackage[super]{nth}
\usepackage{microtype}
\usepackage{tabularx}
\usepackage{supertabular}
\usepackage{physics}
\usepackage[math]{cellspace}

\usepackage{hyperref}
\usepackage{cleveref} % enables \cref, \Cref for referencing multiple figures in one

\title[Shan--Chen interacting vacuum cosmology]{Shan--Chen interacting vacuum cosmology}

\author[N. B. Hogg \& M. Bruni]{
Natalie B. Hogg$^{1, 2}$\thanks{E-mail: natalie.hogg@uam.es}
and Marco Bruni$^{2, 3}$\thanks{E-mail: marco.bruni@port.ac.uk}
\\
$^{1}$Instituto de F\'{i}sica Te\'{o}rica UAM/CSIC, C/Nicol\'{a}s Cabrera 13-15,  Universidad Aut\'{o}noma de Madrid, Cantoblanco, Madrid 28049, Spain\\
$^{2}$Institute  of  Cosmology  and  Gravitation,  University  of  Portsmouth, Burnaby  Road,  Portsmouth  PO1  3FX, United Kingdom\\
$^{3}$INFN Sezione di Trieste, Via Valerio 2, 34127 Trieste, Italy
}

\date{Accepted 2022 February 2. Received 2022 February 2; in original form 2021 September 23}

\pubyear{2022}

\begin{document}
\label{firstpage}
\pagerange{\pageref{firstpage}--\pageref{lastpage}}
\maketitle

\begin{abstract}
In this paper, we introduce a novel class of  interacting vacuum models, based on recasting the equation of state originally developed in the context of lattice kinetic theory by Shan \& Chen as  the  coupling between the vacuum and cold dark matter (CDM). This coupling allows the vacuum to evolve and is nonlinear around a characteristic energy scale $\rho_*$, changing into a linear coupling with a typical power law evolution at scales much lower and much  higher than $\rho_*$. Focusing on the simplest sub-class of models where the interaction consists only of an energy exchange and the CDM remains geodesic, we first illustrate the various possible models that can arise from the Shan--Chen coupling, with several different behaviours at both early and late times depending on the values of the model parameters selected.  We then place the first observational constraints on this Shan--Chen interacting vacuum scenario, performing an MCMC analysis to find those values of the model and cosmological parameters which are favoured by observational data. We focus on models where the nonlinearity of the coupling is relevant at late times, choosing for the reference energy scale $\rho_*$, the critical energy density in $\Lambda$CDM. We show that the observational data we use are compatible with a wide range of models which result in different cosmologies. However, we also show that $\Lambda$CDM is preferred over all of the Shan--Chen interacting vacuum models that we study, and comment on the inability of these models to relax the $H_0$ and $\sigma_8$ tensions.
\end{abstract}


\section{Introduction} \label{sec:intro}
Reports of the death of $\Lambda$CDM are greatly exaggerated. Since the discovery of the accelerating expansion of spacetime more than twenty years ago \citep{Riess1998, Perlmutter1999}, numerous  observations of our Universe have yielded a standard model of cosmology that has stood up to every challenge. This model, $\Lambda$CDM, posits that the Universe is dominated at late times by a cosmological constant, $\Lambda$, which is responsible for the observed accelerating expansion. Myriad observations of structure formation and galaxy evolution (but not experimental particle physics -- yet) have provided robust evidence for the existence of the other major component of our Universe, cold dark matter (CDM) \citep{Zwicky1933, Eke1998, Sofue:2000jx, Frenk:2012ph, Kunz:2016yqy}. 

Nevertheless, problems with the cosmological constant and its use as the dark energy abound \citep{Martin2012}, leading to many attempts to explain the late time acceleration with an alternative model. Examples of alternative dark energy models range from those in which the vacuum interacts with cold dark matter \citep{Wands:2012vg, Salvatelli2014, Martinelli:2019dau, Hogg:2020rdp}, to quintessence \citep{Amendola1999, Linder2007, Park:2021jmi}, k-essence \citep{Amendariz2001, Putter2007} and yet more esoteric models (see e.g. \cite{Copeland2006} or \cite{DiValentino:2021izs} for a review of dark energy models).

Some alternative models of dark energy to the cosmological constant are also motivated by their potential ability to reconcile the more than $4\sigma$ discrepancy between the value of the Hubble parameter at redshift zero ($H_0$) determined from cosmic microwave background (CMB) temperature and polarisation measurements (see e.g. \cite{Aghanim:2018eyx,DiValentino:2020zio}) and that measured using the low redshift standardisable candles Type Ia supernovae (SNIa) calibrated using Cepheid variable stars by the S$H_0$ES collaboration (see e.g. \cite{Riess:2019cxk}). We note that when comparing these two measurements alone, it is preferable to refer to this as a ``supernova absolute magnitude tension'' rather than the more commonly used ``$H_0$ tension'' -- see \cite{Camarena:2021jlr, Efstathiou:2021ocp} for more details on this point. For a different perspective on the discrepancy, see \cite{Haridasu:2020pms}.

However, the S$H_0$ES SNIa are not the only local probe of $H_0$. Taking into account the numerous complementary measurements of $H_0$ now available, such as those from water masers \citep{Pesce:2020xfe}, SNIa calibrated using stars at the tip of the red giant branch of the Hertzsprung--Russell diagram \citep{Sakai1999, Freedman:2019jwv} and strong lensing time delays \citep{Wong:2019kwg}, the familiar tension between local determinations of $H_0$ and measurements from the CMB which rely on a cosmological model  remains. The ``$H_0$ tension'' is thus rightly used by many authors as a motivation to study beyond $\Lambda$CDM cosmologies, most commonly exploring different dark energy models. 

In this context, it is worth noting that in single scalar field quintessence models, where the dark energy equation of state is $w \geq -1$, $H_0$ is always smaller relative to the value obtained in a $\Lambda$CDM universe \citep{Banerjee:2020xcn}, implying  that the $H_0$ tension cannot be resolved in these models.

A consideration sometimes made when searching for alternative dark energy models is that the model should preferably have arisen from pre-existing physics, and hence avoid having been overly designed to solve a specific problem. More generally, this aesthetic standpoint can be seen as a facet of the Bayesian school of statistical thought: ``naturalness'', or our distaste for finely tuned models which produce a desired phenomenology, is simply another prior imposed on the model. 

One promising approach along these lines is to consider dark energy and dark matter as a single cosmological fluid, such as a unified dark matter, or UDM, which behaves according to some equation of state. This simple idea has a long history, dating back to the introduction of the Chaplygin gas model by \cite{Kamenshchik:2001cp}, in which the transition from cold dark matter to dark energy domination is achieved using a perfect fluid with a non-standard equation of state.

However, serious objections to the Chaplygin gas model were raised when it was found that unified models of this type result in oscillations or an exponential blow-up in the matter power spectrum, thus ruling out the vast majority of the viable Chaplygin gas model space \citep{Sandvik:2002jz}. To counter this, further models in which dark matter transitions to dark energy via a condensation mechanism have been proposed and studied, and interest in unified models of this type remains high -- see, for example, \cite{Balbi:2007mz,Pietrobon:2008js,Piattella:2009kt, Piattella:2009da, Gao:2009me, Bertacca:2010mt,DeFelice:2012vd, Wang:2013qy, Li:2018hhy, Li:2019umg}. A review of fluid dark energy models can be found in \cite{Bamba:2012cp}.

The Shan--Chen equation of state was first proposed by Shan and Chen in the context of lattice kinetic theory in \cite{Shan1993}. A fluid with this equation of state behaves as an ideal gas in the low and high density regimes, but has a liquid--gas coexistence curve i.e.\ a region of temperature and pressure in which the fluid can be in both the liquid and gas state. Such an equation of state means that a phase transition can easily arise. 

This model was successfully applied to the cosmological context by \cite{Bini:2014pmk, Bini:2016wqr}, with the finding that a dark energy fluid with a Shan--Chen equation of state naturally evolves towards a Universe with a late time accelerating expansion without the presence of a cosmological constant. In those works, modifications made to the background expansion by the Shan--Chen model were studied, and quantities such as the distance modulus in the Shan--Chen model were compared to Type Ia supernova data.

In this paper, we cast the Shan--Chen equation of state into the form of a coupling between the vacuum and cold dark matter. After studying this Shan--Chen coupling in detail, we implement the interacting model in the public codes \texttt{CAMB} \citep{Lewis:1999bs, Howlett:2012mh} and \texttt{CosmoMC} \citep{Lewis:2002ah, Lewis:2013hha} in order to obtain constraints on the cosmological parameters and the model parameters in the Shan--Chen interacting vacuum model. 

The structure of this paper is as follows: in Section \ref{sec:theory} we outline the theory of the Shan--Chen interacting vacuum model, in Section \ref{sec:method} we describe our methodology, in Section \ref{sec:results} we present the results of our analysis, and in Section \ref{sec:summary} we discuss our results and conclude the paper.

Throughout this work, we assume General Relativity, use units in which the speed of light $c=1$, and assume a spatially flat Friedmann--Lema\^{i}tre--Robertson--Walker (FLRW) background spacetime. 

\section{Interacting vacuum Shan--Chen model} \label{sec:theory}
\subsection{Preliminary: the Shan--Chen equation of state}

The nonlinear equation of state proposed in \cite{Shan1993} is
\begin{align}
P &= w \rho_* \left[\frac{\rho}{\rho_*} + \frac{g}{2} \psi^2\right], \label{eq:ogsc}\\
\intertext{where $\rho$ is the energy density of the fluid, and}
\psi &= 1 - e^{- \alpha \frac{\rho}{\rho_*}},
\end{align}
where $\rho_*$ is a characteristic energy scale and $w$, $g$ and $\alpha$ are free (dimensionless) parameters of the model.

In the Shan--Chen model of dark energy introduced by \cite{Bini:2014pmk,Bini:2016wqr}, the matter--energy content of the Universe is assumed to be a perfect fluid which obeys the equation of state \eqref{eq:ogsc}, and $\rho_*$ is identified with the critical density today: $\rho_*=\rho_{\mathrm{crit}, 0} = \frac{3H_0^2}{8\pi G}$, where  $H_0$ is the value of the Hubble parameter at  redshift zero.

We note that another choice for this energy scale could be, for example, the value of the matter density at matter--radiation equality, as this could result in an early dark energy-type behaviour. An example of a successful early dark energy model can be found in \cite{Poulin:2018cxd}, although that model's competitiveness with $\Lambda$CDM and its ability to resolve the $H_0$ tension was called into question in \cite{Hill:2020osr}, where the analysis of \cite{Poulin:2018cxd} was updated using Planck 2018 data as well as additional information from large scale structure. There have since been a number of other works exploring different possibilities under the early dark energy umbrella, including New Early Dark Energy \citep{Niedermann:2020dwg}, Chain Early Dark Energy \citep{Freese:2021rjq} and Chameleon Early Dark Energy \citep{Karwal:2021vpk}. We leave the exploration of the Shan--Chen dark energy scenario, including early dark energy models, to a future work.

While we do not need to go into the details of the Shan--Chen  dark energy model here, it is useful to point out few features for the sake of comparison with the Shan--Chen interacting vacuum model we propose in the next section. 

To this end, let us consider the energy conservation equation for a Shan--Chen dark energy in an FLRW universe,
\begin{align}\label{ECE}
    \dot{\rho} = -3H(\rho + P(\rho)),
\end{align}
where $P(\rho)$ is given by \eqref{eq:ogsc} and as usual
 $H = \frac{\dot{a}}{a}$ is the Hubble expansion scalar, with $a$ being the cosmic scale factor.

We can then define the effective equation of state parameter $w_\mathrm{eff}$,
\begin{align}
    w_\mathrm{eff}=\frac{P(\rho)}{\rho}=w+\frac{wg\rho_*}{2\rho}(1-e^{-\alpha \frac{\rho}{\rho_*}})^2\;.\label{weff}
\end{align}
Clearly, this only depends on $x=\rho/\rho_*$. Although the energy scale $\rho_*$ is physically meaningful, in that it determines at which energies the nonlinear term in \eqref{weff} becomes relevant, it has no influence on the qualitative behaviour of the Shan--Chen effective equation of state.
 
Depending on the values of the model parameters $\alpha$, $g$ and $w$, there may be stationary points $\rho=\rho_\mathrm{S}$ such that $P(\rho_\mathrm{S})=-\rho_\mathrm{S}$, i.e.\ $w_\mathrm{eff}=-1$ and $\dot{\rho}=0$, from \eqref{ECE}. If they exist, in general they appear in pairs\footnote{There can be none or two, or one when the two coincide.}: they represent cosmological constants, one stable and one unstable, and the dark energy density can either evolve asymptotically toward the stable one, away from the unstable one, or in between the two. 

Around $\rho_*$ the fluid is not ideal, while in the limits $\rho\gg \rho_*$ and $\rho\ll\rho_*$ we get $P=w\rho$ and the usual linear  equation of state is recovered. It is easy to show that $w_\mathrm{eff}$ has a minimum when $wg<0$ and a maximum when $wg>0$; in addition, the position of the minimum or maximum, given by a specific value of $x=\rho/\rho_*$, does not depend on $w$ and $g$, but is shifted by $\alpha$. 
 
In the top left panel of Figure \ref{fig:wint}, we show four examples of  the effect of different parameter values on the behaviour of $w_\mathrm{eff}$ as a function of $x=\rho/\rho_*$:  $w=\pm1/3$ and $g=\pm 8$, keeping a fixed value of $\alpha=2.7$, the best fit value of this parameter found in \cite{Bini:2014pmk} in an analysis of SNIa distance modulus data. Note that there are six general cases for $wg<0$ and six cases for $wg>0$, plus limiting cases. We have chosen to show four examples only to ensure clarity in the plot. We discuss them in more depth in Section \ref{sec:SCint} for easier comparison with the interacting Shan--Chen scenario, but we will give a brief overview here. 

Since  $w_\mathrm{eff}$ is asymptotic to $w$, changing the value of $w$ whilst keeping the other parameters fixed simply shifts the curves up and down, whilst changing the value of $g$ makes the ``bump'' in the curves more or less pronounced. In general, the curves may or may not cross $w_\mathrm{eff}=-1/3$ and $w_\mathrm{eff}=-1$. We can infer from this plot that whether or not acceleration occurs in the Shan--Chen dark energy model, and when, depends on the values of $w$, $g$ and $\alpha$. Note that for the purposes of this plot, and for simplicity's sake, we are assuming a dark energy dominated Universe here, hence acceleration is achieved when $w_\mathrm{eff}<-1/3$.

In general, several different dark energy models can be obtained from \eqref{weff}. For instance, in some cases $w_\mathrm{eff}<-1$, and so the Shan--Chen dark energy can have phantom behaviour, evolving between the two cosmological constants, i.e.\ the two values  $\rho_\mathrm{S}/\rho_*$ for which $w_\mathrm{eff}=-1$. When $w>-1$ and $wg<0$ there are models that would produce acceleration only below a certain energy density threshold, with the energy density decreasing, eventually tending to a cosmological constant when $w_\mathrm{eff}=-1$ is asymptotically approached. We leave a complete analysis of the Shan--Chen dark energy scenario to future work, and now move to a discussion of the Shan--Chen  interacting vacuum model.
  
\subsection{The interacting vacuum scenario}
The classical vacuum is  identified as a perfect fluid with equation of state $P=-V$, where $V$ is the vacuum energy density. If non-interacting, $V$ is equivalent to a cosmological constant. If interacting,  however, $V$ can vary, as we are  now going to summarise. More details can be found in \cite{Wands:2012vg}.

We consider the specific case of the vacuum interacting with  cold dark matter (CDM), as in previous works \citep{Salvatelli2014,Martinelli:2019dau, Hogg:2020rdp}. The vacuum energy--momentum tensor is  
\begin{align}
    \check{T}^\mu_\nu = -V g^\mu_\nu, \label{eq:vacuum_emt}
\end{align}
where $V$ is  the vacuum energy density and $g^\mu_\nu$ is the metric tensor. By comparing \eqref{eq:vacuum_emt} with the  energy--momentum tensor  of a perfect fluid,
\begin{align}
T^\mu_\nu = P g^\mu_\nu + (\rho + P) u^\mu u_\nu, \label{eq:cdm_emt}
\end{align}
where $P$ is the pressure, $\rho$ the energy density and $u^\mu$ the 4-velocity of the fluid, we can identify $V =-P = \rho$.

It is clear from (\ref{eq:vacuum_emt}) that any timelike 4-vector is an eigenvector of $ \check{T}^\mu_\nu$, with $V$ its eigenvalue: thus, $V$ is the energy density of vacuum in any frame, i.e.\ for any observer. 

CDM is represented by a pressureless perfect fluid, hence its energy--momentum tensor is given by
\begin{align}
    T^\mu_\nu = \rho_c u^\mu u_\nu,
\end{align}
where $\rho_c$ is the rest-frame energy density of CDM and $u^\mu$ its 4-velocity. We can  introduce an energy exchange between the two components via an energy--momentum flow 4-vector $Q^\nu$,
\begin{align}
    \nabla_\mu T^\mu_\nu &= - Q_\nu,\label{eq:cdm_cons} \\ 
    \nabla_\mu \check{T}^\mu_\nu &= -\nabla_\nu V = Q_\nu, \label{eq:vacuum_cons}
\end{align}
so that the total energy--momentum tensor is conserved, as it should be in the framework of General Relativity as a consequence of the Bianchi identities. 
This interaction 4-vector can be projected in two parts, one parallel and one orthogonal to the cold dark matter 4-velocity $u^\mu$,
\begin{align}
    Q^\mu = Q u^\mu + f^\mu,
\end{align}
where $Q=-Q_\mu u^\mu$ represents the energy exchange and $f^\mu$ the momentum exchange between cold dark matter and the vacuum, in the frame comoving with CDM, and $f_\mu u^\mu=0$. 

As in previous works \citep{Salvatelli2014,Martinelli:2019dau, Hogg:2020rdp}, we now impose the \textit{geodesic condition} on the interaction, meaning that we neglect the momentum exchange by setting $f^\mu = 0$. Since $f^\mu = \rho_c a^\mu$, where $a^\mu$ is the 4-acceleration, this means that there is no additional acceleration on the cold dark matter particles due to the interaction, and they hence remain geodesic. With this choice, in the synchronous comoving gauge used in \texttt{CAMB}, the interaction is unperturbed and fully encoded in the background $Q$. However, the interaction does enter the evolution equation for  the CDM density contrast. A detailed discussion of the linear perturbations in the interacting vacuum can be found in \cite{Wands:2012vg, Wang:2013qy,Martinelli:2019dau}.

In an FLRW background, \eqref{eq:cdm_cons} and \eqref{eq:vacuum_cons} reduce to
\begin{align}\label{mattev}
    \dot{\rho_c} + 3H \rho_c &= -Q,\\
    \dot{V} &= Q, 
\end{align}
and the Friedmann--Raychaudhuri equation for CDM interacting with vacuum is formally the same as in $\Lambda$CDM:
\begin{align}\label{eq:Ray}
    \frac{\ddot{a}}{a}=-\frac{4\pi G}{3}(\rho_c -2V) ,
\end{align}
where in the cosmological constant case, $V=\Lambda/8\pi G$. Accordingly,  an interacting (and thus evolving) $V>0$ always contributes positively to $\ddot{a}$, eventually accelerating the expansion if and when it becomes the dominant component. This is unlike the case of a dark energy component with an equation of state $P=P(\rho)$ such as the Shan--Chen dark energy model given in \eqref{eq:ogsc} which, depending on its parameter values, may or may not positively contribute to $\ddot{a}$.

\subsection{The Shan--Chen interacting vacuum model}\label{sec:SCint}
We can now recast the Shan--Chen model as a parameterisation of the coupling $Q$ between the vacuum and cold dark matter, thus introducing the Shan--Chen interacting vacuum model. 

Starting from \eqref{ECE}, we substitute  $P$ from \eqref{eq:ogsc}, replace $\rho$ with the vacuum energy density $V$ and, to avoid confusion with the dark energy case, we rename the parameter $w$ as $\beta$. We also introduce the dimensionless parameter $q$ which controls the overall strength of the interaction. This yields the final form of the coupling between the vacuum and cold dark matter,
\begin{align}
Q = \dot{V} =-3H q\left[(1+\beta)V+ \frac{\beta g}{2}  \rho_*  \left(1 - e^{- \alpha \frac{V}{\rho_*}}\right)^2\right]. \label{eq:finalform}
\end{align}
When $q=0$, there is no interaction, and we therefore return to $\Lambda$CDM.

We have a number of additional free parameters in the Shan--Chen interacting vacuum model with respect to $\Lambda$CDM: $q$, $\alpha$, $g$, $\beta$ and $\rho_*$. It is important to note that  identical models can be obtained through a remapping of some of these parameters. Specifically, for any fixed pair $\alpha$ and $\rho_*$, and a triad  $q$,  $\beta$ and $g$, we can choose any $\beta'$ and then choose
\begin{align}
    q' &= \frac{q(1+\beta)}{(1+\beta')}, \\
    g' &= \frac{q \beta g}{q' \beta'},
\end{align}
thus leaving \eqref{eq:finalform} invariant. Therefore, in order to simplify our analysis, we will fix the values of some of these parameters. In the following sections, we will fix the characteristic energy scale to the value of the critical density today in $\Lambda$CDM, i.e. $\rho_*= \bar{\rho}_{\mathrm{crit}, 0}$, as we are interested in the effect that the interaction may have at late times. As previously mentioned, another choice could be the value of the matter density at matter--radiation equality. We will discuss the values chosen for the other parameters below, after showing the effect that changing each one has on the overall behaviour of this interacting vacuum.

Before proceeding to the main analysis, therefore, let us formally define an effective equation of state for the Shan--Chen interacting vacuum, which we call $w_{\mathrm{int}}$, so that we may better understand the behaviour of the vacuum energy density $V$ with this interaction. In analogy with \eqref{ECE} and \eqref{weff} we can rewrite \eqref{eq:finalform} as
\begin{align}\label{eq:Vdot}
Q = \dot{V} =-3H V (1+w_{\mathrm{int}}) ,
\end{align}
where we define\footnote{This specific definition is also motivated by its practical use in \texttt{CAMB}.} 
\begin{equation}
	w_{\mathrm{int}} = q(1+\beta) + \frac{q\beta g}{2x} \left(1- e^{-\alpha x}\right)^2 -1, \label{eq:wint}
\end{equation}   
and we have now defined $x= V/\rho_*$. Clearly, again in analogy with the dark energy case, there can be values $V=V_\mathrm{S}$ such that $w_{\mathrm{int}} =-1$, where $\dot{V}=0$. If they exist, these stationary points appear in pairs and they represent cosmological constants, one stable and one unstable, separating models where $V$ decays from models where $V$ grows. Thus there is again a variety of possible models; for instance, models where $V$ evolves between these two constants, as well as models where $V$ asymptotically  decays or grows  toward the stable cosmological constant (stationary point). 

\begin{figure*}
\includegraphics[width=0.49\textwidth]{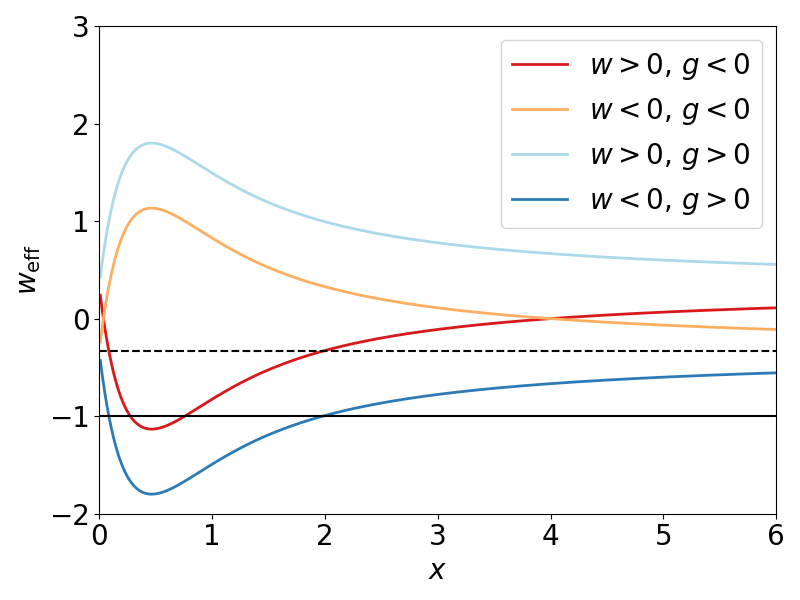}
\includegraphics[width=0.49\textwidth]{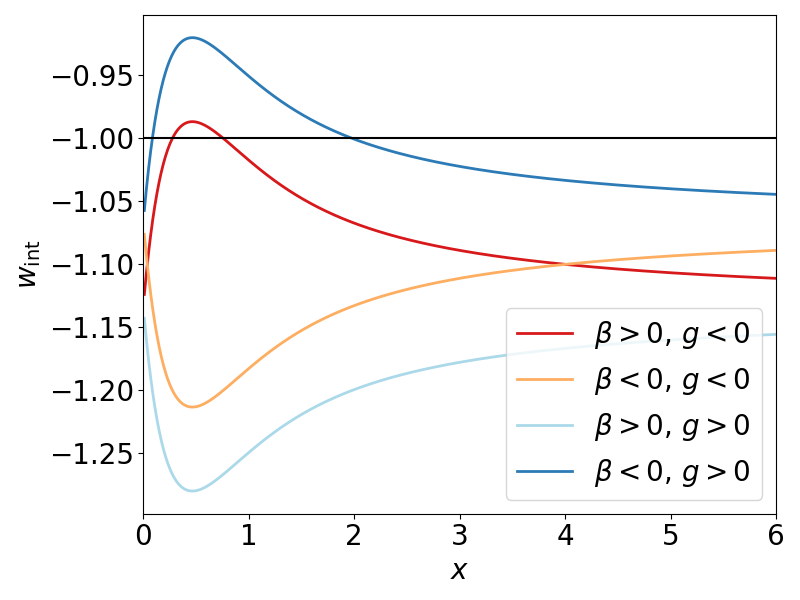} \\
\includegraphics[width=0.49\textwidth]{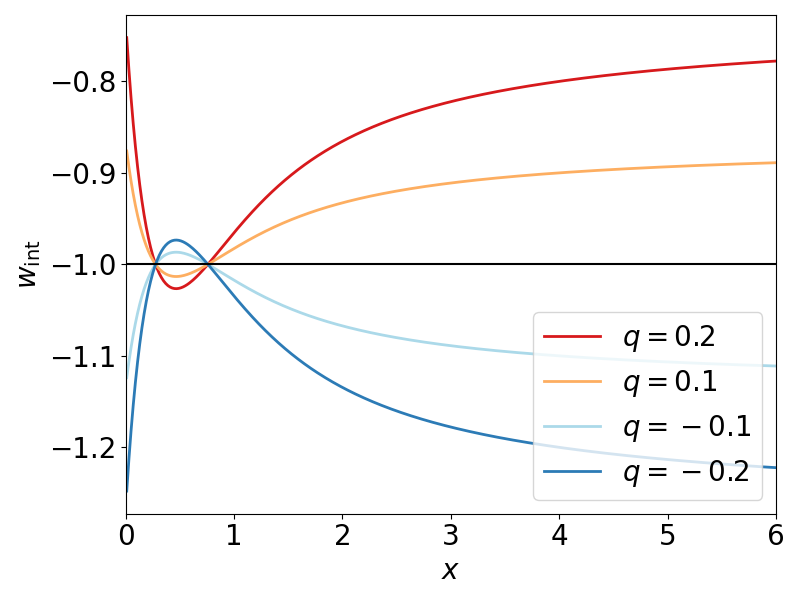}
\includegraphics[width=0.49\textwidth]{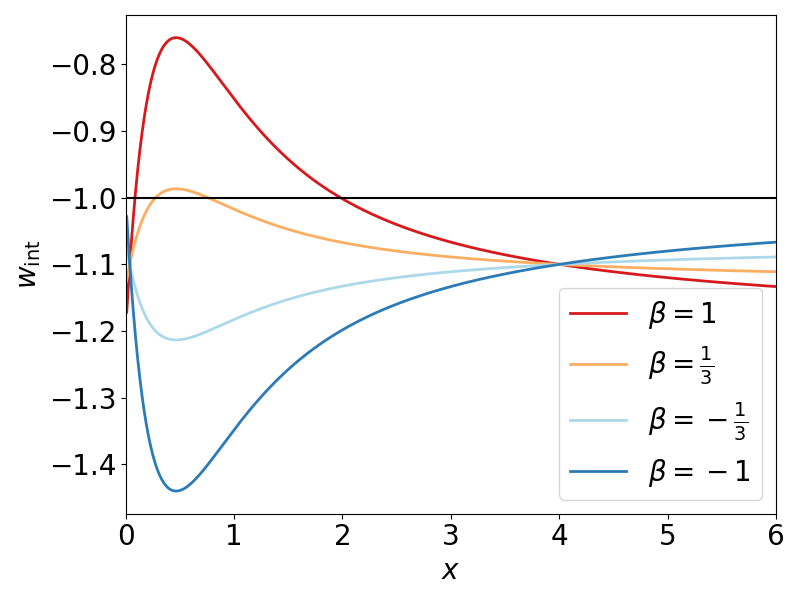}\\
\includegraphics[width=0.49\textwidth]{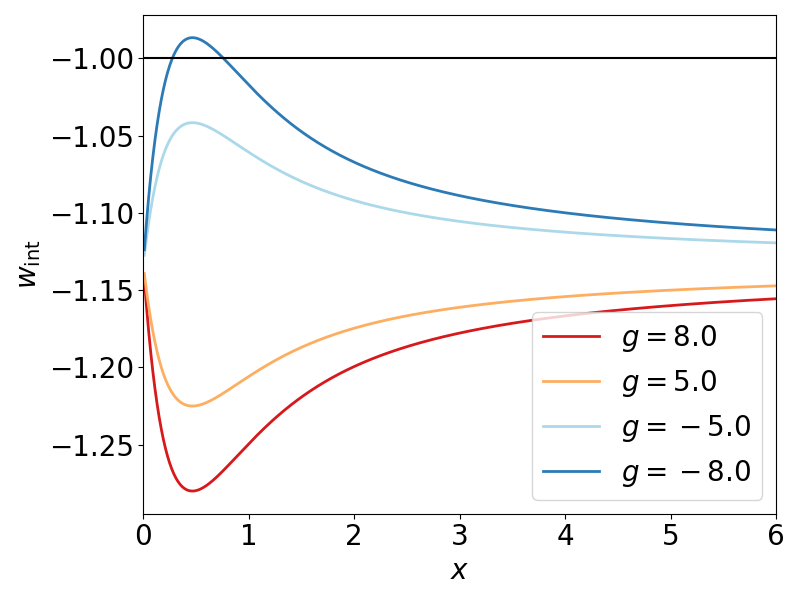}
\includegraphics[width=0.49\textwidth]{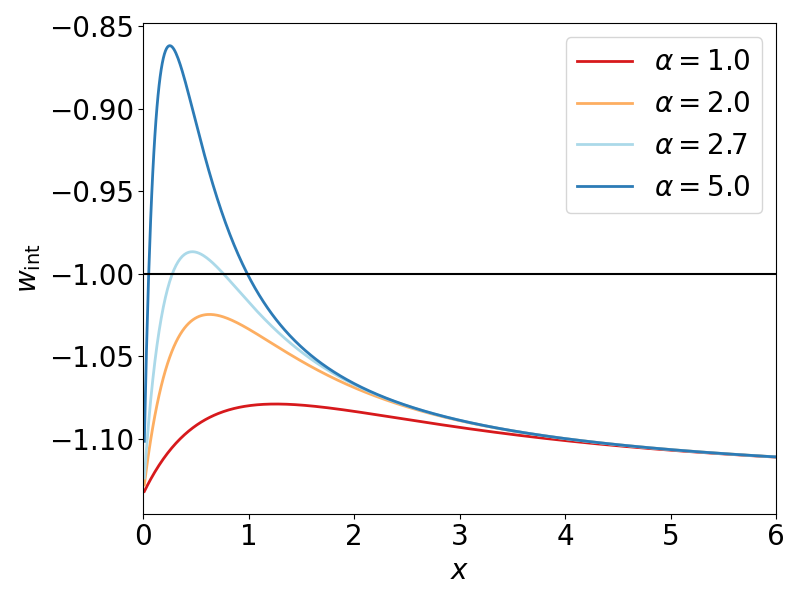}
\caption{Upper panels: four examples of the behaviour of $w_{\rm eff}$ (upper left) and $w_{\rm int}$ for $w=\pm 1/3$ (a.k.a. $\beta$) and $g=\pm 8$ as a function of $x$, the energy density normalised to the energy scale $\rho_*$. In the top-left panel the dashed black line  marks the transition to the accelerating regime for $w_{\rm eff}<-1/3$ and the solid black line  separates phantom models, with $w_{\rm eff} < -1$, from standard dark energy ones. Similarly, in all other panels   the  black solid line separates models where the vacuum energy density $V$ decays, with $w_{\rm int} > -1$, from models where $V$ grows. In the top-right panel we have chosen a negative $q=-0.1$, so that  curves here have the opposite behaviour of those with the same colour (same parameter values) in the top-left panel. Middle and lower panels: the behaviour of $w_{\rm int}$ as a function of $x=V/\rho_*$ for different values of $q$ (middle left), $\beta$ (middle right), $g$ (lower left) and $\alpha$ (lower right). When not changed, the parameters are kept fixed to $q=-0.1$, $\beta=1/3$, $g=-8$ and $\alpha=2.7$. 
} \label{fig:wint}
\end{figure*}

In Figure \ref{fig:wint}, we show the effects of different parameter values on the behaviour of the function $w_{\mathrm{int}}$, comparing it with the non-interacting equation of state $w_{\mathrm{eff}}$ \eqref{weff} of the original Shan--Chen fluid dark energy models in the upper panels. 

In the upper left panel we show four examples for $w_{\mathrm{eff}}$, given by $w=\pm 1/3$ and $g=\pm 8$, with $\alpha$ fixed to the best fit value of $\alpha=2.7$ found by \cite{Bini:2014pmk}. We can see that  when $wg>0$, i.e.\ the light blue and yellow curves, the effective equation of state is always outside the accelerating regime, i.e.\ always $w_{\rm eff}> -1/3$. However, when $wg<0$, i.e.\ the red and dark blue curves, the effective equation of state can allow for an accelerating phase. Note that in all the plots in this figure, values of $x\gg 1$ correspond to high energies, i.e.\ energies which are large with respect to the reference energy scale $\rho_*$.

For the case where $w$ is positive and $g$ is negative (red curve, $w=1/3$), the Shan--Chen  fluid behaves  as radiation at high energies in the region $w_{\mathrm{eff}}>-1$, decaying and transitioning through a dark energy accelerating phase and toward a cosmological constant (the $x$ value corresponding to the right-most point on the line $w_{\mathrm{eff}}=-1$ crossed by the red line). 

On the other hand, in the region $w_{\mathrm{eff}}<-1$,  the Shan--Chen  fluid is trapped in a phantom regime, growing between the two cosmological constants corresponding to  the $x$ values where the line $w_{\mathrm{eff}}=-1$ is crossed by the red line. When $w\leq -1/3$  and $g$ is positive (dark blue curve), the Shan--Chen fluid behaves as a dark energy even at high energies, again evolving toward a cosmological constant;  the phantom regime is qualitatively the same as for the red curve. 

In the upper right panel of Figure \ref{fig:wint}, we show the same cases (the same parameter values corresponding to the same colours)  for the Shan--Chen interacting  vacuum, i.e.\ the effective equation of state $w_{\mathrm{int}}$ given in \eqref{eq:wint}, fixing $q=-0.1$. Note that the y-axis has a different scale to the y-axis in the upper left panel and, because $q<0$, the behaviour of curves with the same colour is opposite. However, as discussed after \eqref{eq:Ray}, in the interacting vacuum case $V$ always gives a positive contribution to  acceleration; therefore for the Shan--Chen interacting vacuum this is independent from the sign combination of $\beta$ and $g$ (recalling that we renamed $w$ as $\beta$ in the interacting case). For  $w_{\mathrm{int}}<-1$ there can be  cases allowing for $V$ growing between two cosmological constants (not shown),  corresponding  to the dark energy phantom cases in the top left panel, as well as cases where $V$ grows indefinitely from zero (yellow and light blue lines) or from a cosmological constant (blue and red lines). Above $w_{\mathrm{int}}=-1$ the red and blue lines give models where $V$ decays between the two cosmological constants. 

In the rest of Figure \ref{fig:wint}, we show the effects on $w_{\rm int}$ of different values of $q$ (middle left panel), different values of $\beta$ (middle right panel), different values of $g$ (bottom left panel) and different values of $\alpha$ (bottom right). If the parameter in question is not shown with different values it is kept fixed to the best fit values of \cite{Bini:2014pmk}: $\beta=1/3$, $g=-8.0$ and $\alpha=2.7$, with $q$ fixed to $-0.1$ as an example.

Starting with the middle left panel of Figure \ref{fig:wint}, we can see the effect of changing both the magnitude and sign of the interaction strength $q$; in particular, changing the sign of $q$ simply mirrors curves with same parameter values about the $w_{\rm int} = -1$ line. In all cases shown there are two cosmological constants where the curves cross the $w_{\rm int} = -1$ line. In general, at high energies (i.e. $x\gg 1$) $w_{\rm int}$ behaves as a constant, then the nonlinear term in \eqref{eq:wint} becomes dominant around $x=1$. For models above the $w_{\rm int} = -1$ line  (red and yellow lines) $V$ decays from high energies to the cosmological constant on the right, or to zero from the cosmological constant on the left. For models below the $w_{\rm int} = -1$ line  (light blue and dark blue lines) $V$ grows  from  the cosmological constant on the right to high energies, or from zero to the cosmological constant on the left. Models in between the two cosmological constants decay or grow between the two for $w_{\rm int} > -1$ and $w_{\rm int} < -1$, respectively.

Next, in the middle right panel of Figure \ref{fig:wint}, we see the effect that changing $\beta$ has. Generally, increasing $\beta$ moves the curves up ($w_{\rm int} =q(1+\beta) -1$ at $x=0$) and increases the relevance of the nonlinear term in \eqref{eq:wint} at low energies while shifting the point at which this term becomes dominant over the constant one to higher energies. In addition, changing the sign of $\beta$ flips the curves upside down around the line  $w_{\rm int} =q -1$, as is also clear from the red and the yellow curves in the upper right panel.

Third, in the lower left panel of Figure \ref{fig:wint}, we can see the effect of changing $g$. The result is similar to that of $\beta$, with no vertical translation. In addition, changing the sign of $g$ flips the curves upside down around  the line  $w_{\rm int} =q(1+\beta) -1$.

Finally, in the lower right panel of Figure \ref{fig:wint}, we can see the effect of changing $\alpha$. This parameter is called the saturation scale by \cite{Bini:2014pmk}, as in the fluid Shan--Chen dark energy model it represents the typical density above which $\psi$ undergoes a saturation effect, i.e. $\psi \approx 1$. For this reason, $\alpha$ should always be positive. We can see that changing $\alpha$ greatly affects the relevance and steepness of the exponential term in  $w_{\rm int}$, while at the same time shifting the $x$ position of the extrema points.

\begin{figure}
	\includegraphics[width=\columnwidth]{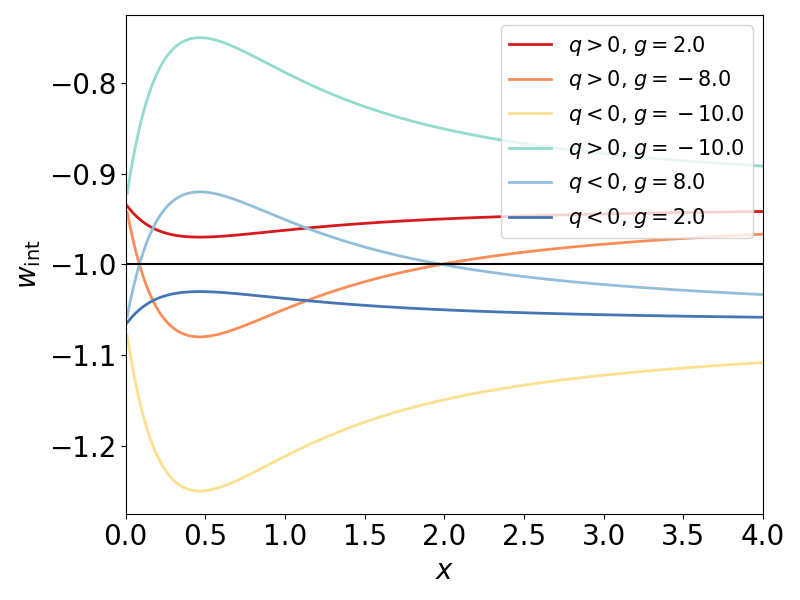}
	\caption{The six possible cases of $w_{\rm int}$, for a total of ten possible  evolution models for  the vacuum energy density $V$, decreasing for $w_{\rm int}>-1$ and increasing for $w_{\rm int}<-1$. The cyan, red, blue and yellow lines each represent a model where $V$ varies between zero and infinity, either growing or decaying in time. Each of the pale blue and orange lines represent three different models, separated by the two points at the crossing of the $w_{\rm int}=-1$ line. These points represent cosmological constants, one stable and one unstable, i.e.\ asymptotic states in the evolution of each of the models represented by these lines.
	%Cyan and light blue lines represent models where the vacuum energy density $V$ decays to zero, yellow and blue lines represent models where $V$ increases from zero. 
	}
	\label{fig:wint_minmax}
\end{figure}

In Figure \ref{fig:wint_minmax} we summarise all the qualitatively different models that can arise in the Shan--Chen  interacting vacuum scenario, depending on the behaviour of $w_{\rm int}(x)$.  First of all, we only illustrate cases where $w_{\rm int}<0$ always; otherwise vacuum may never become dominant over CDM, or not dominant enough, or for the right period of time. Secondly, vacuum decreases or increases for models above or below the $w_{\rm int}=-1$ line respectively, representing $\Lambda$CDM. Lastly, the asymptotic value of $w_{\rm int}$ for large $x$ is the same as $w_{\rm int}(0)$, i.e.\ $w_{\rm int}=q(1+\beta)-1$, and $w_{\rm int}$ has a single minimum or maximum in between. Thus the curve $w_{\rm int}(x)$ can either be entirely above or below $w_{\rm int}=-1$, or $w_{\rm int}=-1$ can be crossed at two points. Hence, there are six possible different curves. For each of the two cases crossing $w_{\rm int}=-1$ there are three models. Otherwise, each curve represents a model; there are thus ten different types of model in total.  

The two cases where $w_{\rm int}>-1$ for any $x$, i.e. the red and cyan lines,  represent two models where  vacuum decays to zero in the far future, coming from a power law decay $V \sim a^{-3q(1+\beta)}$ ($q>0, \beta>-1$) in the far past, at high energies, for $x\gg 1$. Conversely, the two cases in which $w_{\rm int}<-1$ for any $x$, i.e. the dark blue and yellow  lines,  represent two models where  vacuum grows from  zero in the past, eventually growing as the power law  $V \sim a^{-3q(1+\beta)}$ ($q<0, \beta>-1$) in the far future, for $x\gg 1$. 

Finally, each line that is the same type as the pale blue and orange lines represents three models. This is because the two points where these lines cross $w_{\rm int}=-1$ represent stationary points of \eqref{eq:Vdot}, i.e. cosmological constants, that cannot be crossed by the $V$ evolution; they are asymptotic values, either in the past or in the future. Thus, the portion of the orange line at large $x$ represents a model where $V$ decays, initially as  $V \sim a^{-3q(1+\beta)}$ and eventually tending to a constant. The portion of the pale blue line at large $x$ represents a model where in the past  $V$ is asymptotic to a constant, then grows in the future as $V \sim a^{-3q(1+\beta)}$. The portion of the orange line below $w_{\rm int}=-1$ represents a model where $V$ grows between the two cosmological constants, while the portion of the pale blue line above $w_{\rm int}=-1$  represents a model where $V$ decays,  from a constant value in the past to a smaller constant value in the future. 

The portion of the orange and pale blue lines between $x=0$ and the point where $w_{\rm int}=-1$ is crossed represent models where $V$ either decays from zero to a cosmological constant at $x<1$ or vice versa. These type of models are not relevant here, where we assume $\rho_*=\rho_{{\rm crit},0}$, but could be relevant when studying the case of a higher energy $\rho_*$. For example, models represented by this portion of the orange line, with $V$ decaying from a past cosmological constant as a power law  $V \sim a^{-3q(1+\beta)}$, could be relevant when using the Shan--Chen interacting vacuum to mimic the type of early dark energy model presented in \cite{Poulin:2018cxd}, where the dark energy follows precisely this type of evolution. Finally, models represented by this portion of the pale blue line could be used to study the case of an emerging vacuum, which initially grows from zero as $V \sim a^{-3q(1+\beta)}$ and asymptotically tends to a cosmological constant.

Having demonstrated the effect of all the model parameters on $w_{\rm int}$, we now choose to reduce our parameter space somewhat by fixing two of them. For the remainder of our analysis, we fix $g=-8.0$ and $\alpha=2.7$, both of which are the best fit values found by \cite{Bini:2014pmk}. We fix these two parameters in particular, as changing their values is rather similar to the effect of $\beta$, and would result  in a degeneracy with $\beta$ when sampling.

\begin{figure}
	\includegraphics[width=\columnwidth]{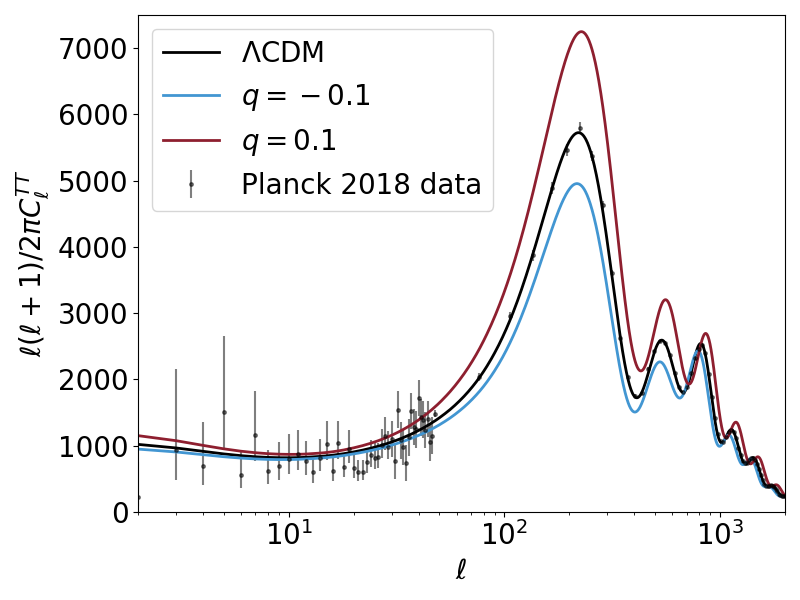} \\
	\includegraphics[width=\columnwidth]{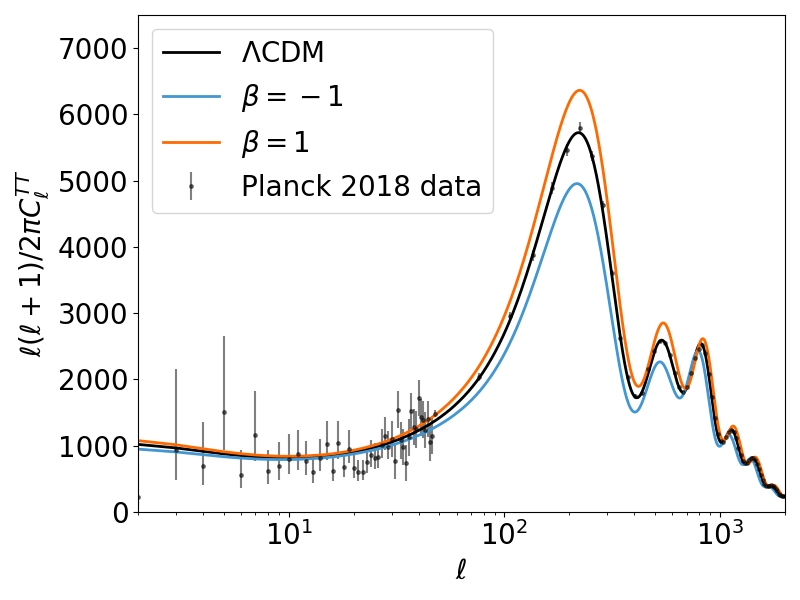}
	\caption{The CMB temperature--temperature power spectra for different values of $q$ (upper panel) and different values of $\beta$ (lower panel). For the upper panel, $\beta$ is kept fixed to $-1$ and for the lower panel, $q$ is kept fixed to $-0.1$, hence the light blue curves in both plots represent the same case.}
	\label{fig:tt}	
\end{figure}

\begin{figure}
	\includegraphics[width=\columnwidth]{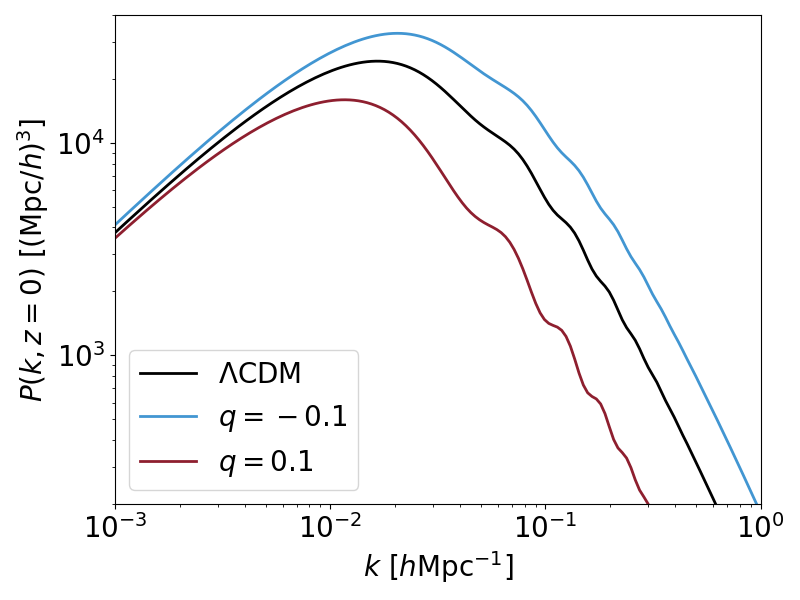} \\
	\includegraphics[width=\columnwidth]{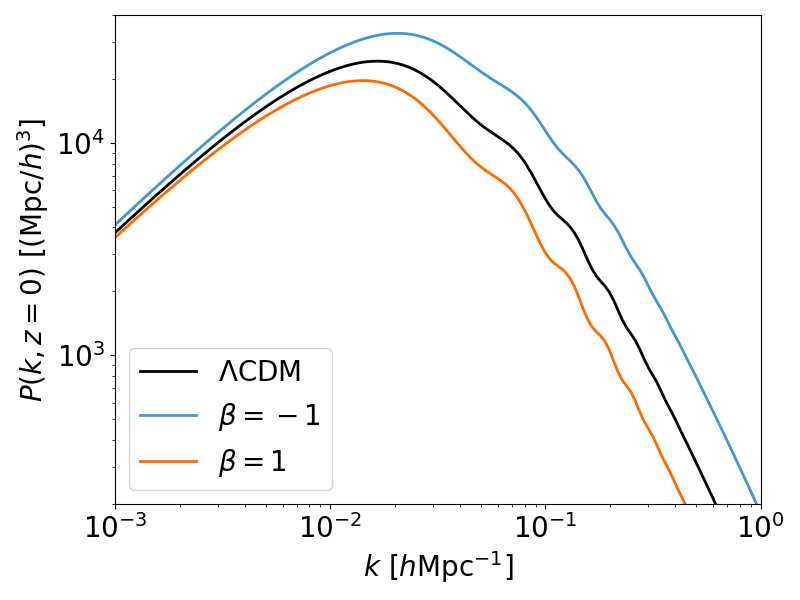}
	\caption{The matter power spectrum at $z=0$ for different values of $q$ (upper panel) and for different values of $\beta$ (lower panel). For the upper panel, $\beta$ is kept fixed to $-1$ and for the lower panel, $q$ is kept fixed to $-0.1$, hence the light blue curves in both sets of plots represent the same case.}
	\label{fig:pk}	
\end{figure}

To further demonstrate that the general effect is still that of an interacting vacuum (as seen in previous works, e.g. \cite{Martinelli:2019dau}), in Figures \ref{fig:tt} and \ref{fig:pk} we plot the CMB temperature--temperature power spectrum and matter power spectrum at $z=0$ for different values of $q$ with $\beta$ fixed,  and for different values of $\beta$ with $q$ fixed. For the purposes of these plots, we keep the cosmological parameters fixed to the Planck 2018 best fits \citep{Aghanim:2018eyx}. As expected, the presence of a coupling between the vacuum and cold dark matter acts to boost or suppress the peaks of the CMB power spectrum, with an opposite effect on the matter power spectrum. These plots also serve as an order of magnitude guide for the priors we will set on these parameters in our MCMC analysis which we describe in the next section.

\section{Method and data} \label{sec:method}
In order to constrain the free parameters of the Shan--Chen interacting vacuum model ($q$ and $\beta$) along with the cosmological parameters, we modify the publicly available Boltzmann code \texttt{CAMB} and its associated MCMC sampler \texttt{CosmoMC}. We implement equation \eqref{eq:finalform} in \texttt{CAMB}, fixing $\rho_*$ to the value of $\bar{\rho}_{{\rm crit}, 0}$, i.e. the critical energy density today in $\Lambda$CDM, $\bar{\rho}_{\rm crit, 0} = 3H_0^2 / 8\pi G = 8.5 \times 10^{-27}$kg m$^{-3}$ when $H_0 = 67.4$ kms$^{-1}$Mpc$^{-1}$ \citep{Aghanim:2018eyx}.

Using \texttt{CosmoMC}, we sample the posterior distributions of the baryon and cold dark matter densities $\Omega_bh^2$ and $\Omega_ch^2$, the amplitude of the primordial power spectrum and the spectral index $A_s$ and $n_s$, the value of the Hubble parameter today, $H_0$, as well as the Shan--Chen parameters $q$ and $\beta$. We keep $\alpha$ fixed to 2.7 and $g$ fixed to $-8.0$. We list the flat priors we impose on the sampled parameters in Table \ref{tab:priors}. 

We use the Planck 2018 measurements of the CMB temperature and polarisation \citep{Aghanim:2018eyx} together with the BAO measurements from the 6dF Galaxy Survey \citep{Beutler2011}, the SDSS Main Galaxy Sample \citep{Ross:2014qpa} and the SDSS DR12 consensus catalogue \citep{Alam:2016hwk}, and the Pantheon catalogue of Type Ia supernovae \citep{Scolnic:2017caz}. 

Note that, in contrast to some previous works, we \textit{do not} include the S$H_0$ES measurement of $H_0$ in our chosen set of data. This is because \textit{it is nonsensical to combine datasets which are in tension in a statistical analysis of this kind}. This statement is hardly original (see e.g. \cite{Hill:2020osr, Ivanov:2020ril}), yet bears repeating due to the proliferation of analyses which claim to resolve the $H_0$ tension while using S$H_0$ES (see some of the results listed in Table B2 of \cite{DiValentino:2021izs}). If the S$H_0$ES result is to be used when constraining an alternative cosmological model, it must first be checked that the value of $H_0$ in that model that has been found by a combination of cosmological-model-dependent datasets, such as CMB+BAO+SNIa, is consistent with the S$H_0$ES value itself. Only then is it safe to run an analysis using e.g. CMB+BAO+SNIa+S$H_0$ES. Furthermore, as recently noted, it may be more appropriate to consider a prior on the supernova absolute magnitude rather than $H_0$ when using the S$H_0$ES data \citep{Benevento:2020fev,Camarena:2021jlr, Efstathiou:2021ocp}

\begin{table}
\centering
\begin{tabular}{SlSl}
	\hline
	\hline
	Parameter          & Prior \\
	\hline
	$\Omega_bh^2$      &  $[0.005,0.1]$\\
	$\Omega_ch^2$      &  $[0.001,0.99]$ \\
	$\log{10^{10}A_s}$ &  $[2.0,4.0]$\\
	$n_s$              &  $[0.8,1.2]$\\
	$H_0$              &  $[50,100]$\\
	$q$                & $[-0.25, 0.25]$\\
	$\beta$            & $[-1.0, 1.0]$ \\
	\hline
\end{tabular}
\caption{Prior ranges of the parameters sampled in our analysis.}\label{tab:priors}
\end{table}

\section{Results and discussion} \label{sec:results}
We now proceed to the results of our MCMC analysis, beginning with case I, in which we keep $\beta$ fixed and sample $q$, followed by case II, in which we keep $q$ fixed and sample $\beta$, and finally case III, where we sample both. Note that the following is not an exhaustive exploration of every parameter combination, but serves to represent the main models under the Shan--Chen interacting vacuum umbrella. For the sake of completeness, we include the study of two additional sub-cases in Appendix \ref{sec:appendix}.

\subsection{Case I: different values of $q$}
In Figure \ref{fig:qfixed}, we show the constraints obtained using the full combination of data: CMB plus BAO plus supernovae. We study a case in which we fix $q$ to zero (i.e. the $\Lambda$CDM limit), which we call case Ia; a case in which we fix $q=0.1$, which we call case Ib; a case in which we fix $q=-0.1$, which we call case Ic; and a case in which we sample over $q$ along with the cosmological parameters, which we call case Id. In all of the sub-cases shown here, $\beta$ is fixed to $1/3$.

We see that for all these cases, even when we sample over $q$ (yellow contours) and therefore have an additional degree of freedom, the resulting cosmologies are constrained by the data to be rather  close to $\Lambda$CDM. We quantify this in Table \ref{tab:results}, where we report the mean posterior values and $1\sigma$ limits obtained using \texttt{GetDist} \citep{Lewis:2019xzd} for this and all subsequent cases studied. The value of $q$ found in this case (Id) is $q=-0.007 \pm 0.1$, which is completely consistent with the $\Lambda$CDM limit of $q=0$.

Since all the present-day values of the cosmological parameters in these models are constrained to be very close to $\Lambda$CDM, we also see no resolution of the tensions present in $\Lambda$CDM. To resolve the $H_0$ tension with this combination of data we would need to see $H_0$ reaching higher values, e.g. $71$ to $73$ kms$^{-1}$ Mpc$^{-1}$. Besides the $H_0$ tension there also exists a lesser tension between the value of $\sigma_8$ obtained from Planck \citep{Aghanim:2018eyx} and from large scale structure surveys such as the Dark Energy Survey (DES) \citep{Abbott:2020knk} and the Kilo Degree Survey (KiDS) \citep{Heymans2020}. A resolution of this tension would require a smaller value of $\sigma_8$ with this combination of data -- again, we do not find this.

\subsection{Case II: different values of $\beta$}
Next, in Figure \ref{fig:w}, we show the result of fixing $q=-0.1$ and fixing $\beta$ to two different values, $1/3$ and $-1/3$, and then sampling it in the range $[-1.0, 1.0]$. In this plot, we see that the posterior distribution of $\beta$ in the case which it is sampled over (purple contours, case IIb) describes the allowed range of the posterior distributions that result from $\beta$ being fixed within the range of the prior used when it is sampled, $\beta \in [-0.1, 0.1]$. The other two cases shown are $\beta=1/3$ (blue contours; note that this case is identical to case Ic) and $\beta=-1/3$ (orange contours, case IIa). The mean posterior values for all these cases are again shown in Table \ref{tab:results}. 

Interestingly, we see that in case IIa, a larger value of $H_0$ is found, demonstrating how interacting vacuum models can in general be invoked to resolve the tension in this parameter. However, the value of $H_0$ in this case does not provide a full resolution of the tension, but merely a slight relaxation, since its marginalised posterior value is $69.2 \pm 0.5$ kms$^{-1}$ Mpc$^{-1}$. We can see from the very large value of $\sigma_8$ that the tension in this parameter is significantly \textit{worsened} in this model. 

\subsection{Case III: sampling $q$ and $\beta$}
\begin{figure}
	\centering
	\includegraphics[width=\columnwidth]{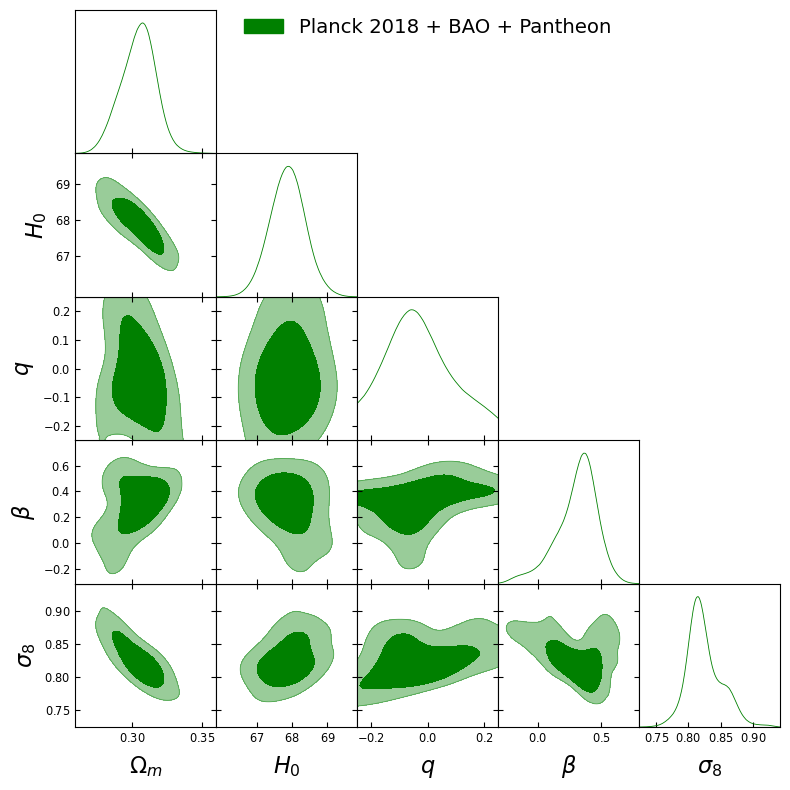}
	\caption{Case III: constraints on $\Omega_{\mathrm{m}}$, $H_0$, $q$, $\beta$ and $\sigma_8$ for the Shan--Chen model with both $q$ and $\beta$ sampled over the ranges $[-0.25, 0.25]$ and $[-1.0, 1.0]$ respectively.}
	\label{fig:qw}
\end{figure}

We finally consider the situation in which we sample over both $q$ and $\beta$. The constraints on the cosmological and model parameters for this case are shown in Figure \ref{fig:qw}. We can see in this case that $q$ is relatively unconstrained but that $\beta$ is constrained to be $0.39 \pm 0.2$. However, we do not necessarily expect $\beta$ to be constrained to be close to zero, as zero is not a $\Lambda$CDM limit of the model; in fact, $\beta = 0$ corresponds to the Cfix case studied in \cite{Martinelli:2019dau}.

A  reason for $q$  to be less constrained  than $\beta$ can be found in the best fit value for $\beta$ that we find. Indeed, for $\beta\approx 0.35$ (and for the given fixed values of $g$ and $\alpha$ that we use)  the today's value of the function $w_{\mathrm int}$ in \eqref{eq:wint}  is very close to $-1$, resulting is a slow evolution of $V$ that is then relatively unaffected by the value of $q$ in the given prior.

\begin{figure}
	\centering
	\includegraphics[width=\columnwidth]{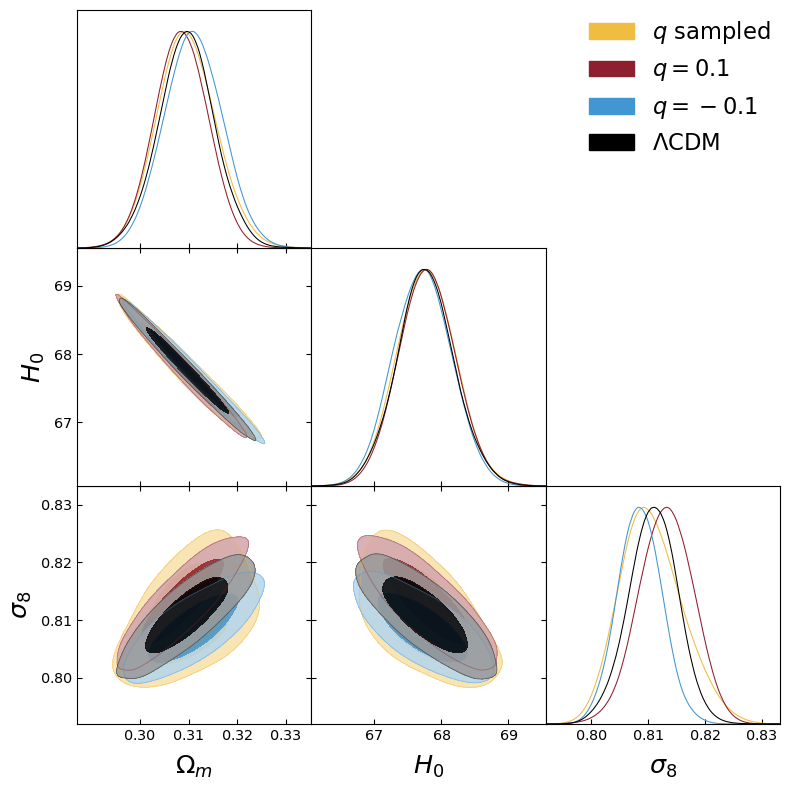}
	\caption{Case I: constraints on $\Omega_{\mathrm{m}}$, $H_0$ and $\sigma_8$ for $\Lambda$CDM (black), and the Shan--Chen model with $q$ fixed to $0.1$ (red), $-0.1$ (blue) and sampled over the range $[-0.25, 0.25]$ (yellow), using the Planck 2018 plus BAO plus Pantheon supernova catalogue. In all of these cases, $\beta=1/3$.}
	\label{fig:qfixed}
\end{figure}

\begin{figure}
\centering
\includegraphics[width=\columnwidth]{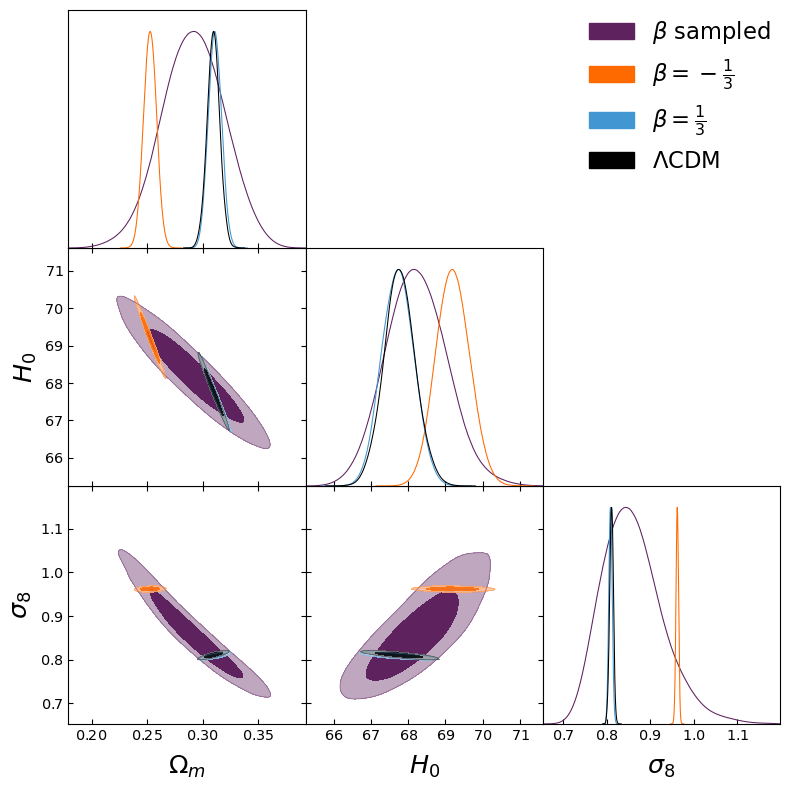}
\caption{Case II: constraints on $\Omega_{\mathrm{m}}$, $H_0$ and $\sigma_8$ for the Shan--Chen model with $q=-0.1$ and $\beta=1/3$ (blue; note that this case is identical to the blue contour shown in Figure \ref{fig:qfixed}), $\beta=-1/3$ (orange) and $\beta$ sampled over the range $[-1.0, 1.0]$ (purple). }
\label{fig:w}
\end{figure}

\begin{table*}
\centering
\begin{tabular}{SlSll|lll} 
	\hline
	\hline
	Parameter        & Case          & Value & Parameter  & Case & Value\\
	\hline
	            & Ia  & $0.02244 \pm 0.00013$ &       & Ia  & $67.8 \pm 0.4$ \\
	            & Ib  & $0.02243 \pm 0.00013$ &       & Ib  & $67.8 \pm 0.4$ \\
	            & Ic  & $0.02244 \pm 0.00013$ &       & Ic  & $67.7 \pm 0.4$ \\
$\Omega_b h^2$	& Id  & $0.02244 \pm 0.00013$ & $H_0$ & Id  & $67.8 \pm 0.4$ \\
	            & IIa & $0.02235 \pm 0.00013$ &       & IIa & $69.2 \pm 0.5$ \\
	            & IIb & $0.02241 \pm 0.00014$ &       & IIb & $68.2 \pm 0.8$ \\
	            & III & $0.02243 \pm 0.00013$ &       & III & $67.9 \pm 0.5$ \\
\hline
	            & Ia  & $0.119 \pm 0.0010$ &            & Ia  & $0.8110 \pm 0.0044$ \\ 
	            & Ib  & $0.119 \pm 0.0010$ &            & Ib  & $0.8132 \pm 0.0048$ \\ 
                & Ic  & $0.120 \pm 0.0010$ &            & Ic  & $0.8086 \pm 0.0039$ \\
$\Omega_c h^2$	& Id  & $0.119 \pm 0.0011$ & $\sigma_8$ & Id  & $0.8107 \pm 0.0055$  \\
	            & IIa & $0.098 \pm 0.0013$ &            & IIa & $0.9619 \pm 0.0030$ \\
	            & IIb & $0.123 \pm 0.0102$ &            & IIb & $0.8594 \pm 0.071$ \\
	            & III & $0.117 \pm 0.0041$ &            & III & $0.8245 \pm 0.028$ \\
\hline
	               & Ia  & $3.050 \pm 0.005$ &     & Ia  &~ $0.0$ \\
	               & Ib  & $3.050 \pm 0.006$ &     & Ib  &~ $0.1$   \\
 	               & Ic  & $3.050 \pm 0.006$ &     & Ic  & $-0.1$    \\
$\log{10^{10}A_s}$ & Id  & $3.050 \pm 0.006$ & $q$ & Id  & $-0.008 \pm 0.1$ \\
	               & IIa & $3.053 \pm 0.006$ &     & IIa & $-0.1$ \\
	               & IIb & $3.051 \pm 0.006$ &     & IIb & $-0.1$\\
	               & III & $3.050 \pm 0.006$ &     & III & $-0.028 \pm 0.1$ \\
\hline
	  & Ia  & $0.9667 \pm 0.0035$ &         & Ia  &~ $1/3$  \\
	  & Ib  & $0.9674 \pm 0.0036$ &         & Ib  &~ $1/3$ \\
	  & Ic  & $0.9675 \pm 0.0036$ &         & Ic  &~ $1/3$\\
$n_s$ & Id  & $0.9675 \pm 0.0036$ & $\beta$ & Id  &~ $1/3$  \\
	  & IIa & $0.9645 \pm 0.0036$ &         & IIa & $-1/3$ \\
	  & IIb & $0.9665 \pm 0.0040$ &	        & IIb &~ $0.12 \pm 0.32$ \\
	  & III & $0.9673 \pm 0.0039$ &         & III &~ $0.31 \pm 0.16$  \\
\hline
\end{tabular}
\caption{Marginalised parameter values with $1\sigma$ errors for each case. Note that all parameters quoted are dimensionless, besides $H_0$ which has dimensions of kms$^{-1}$Mpc$^{-1}$.}\label{tab:results}
\end{table*}

\subsection{Statistical and physical comparison of models} \label{subsec:modelcomparison}
In Table \ref{tab:comparison}, we show the $\chi^2$ and $\Delta \chi^2=\chi^2_{\rm SC}-\chi^2_{\Lambda{\rm CDM}}$ for each case studied. Defined in this way, a negative $\Delta \chi^2$ thus implies that the model under consideration is a better fit to the data than $\Lambda$CDM, apart from in cases Id, IIb and III, where we must compute the significance of the change in $\chi^2$ using a $\chi^2$ difference table (e.g. \cite{chi2book}). This is because these cases have additional degrees of freedom with respect to $\Lambda$CDM. Case Id and IIb each have one additional degree of freedom, because $q$ is allowed to vary in the former and $\beta$ in the latter. Case III have two additional degrees of freedom because both $q$ and $\beta$ are allowed to vary.

From Table \ref{tab:comparison}, we can see that out of the cases with the same number of degrees of freedom as $\Lambda$CDM, we have no cases with a negative $\Delta \chi^2$. This implies that none of these models are a better fit to the data than $\Lambda$CDM. For the comparison of $\Lambda$CDM with cases with additional degrees of freedom, we choose a significance level of 95\%. This means that for one additional degree of freedom, a $|\Delta \chi^2| > 3.841$ would be considered significant. The $\Delta \chi^2$ values for case Id and IIb are both smaller than this value, so $\Lambda$CDM is a better fit than these models. For two additional degrees of freedom, a $|\Delta \chi^2| > 5.991$ is significant. The $\Delta \chi^2$ for case III is $0.82$, so $\Lambda$CDM is again a better description of the data than this model.

\begin{figure}
    \centering
    \includegraphics[width=\columnwidth]{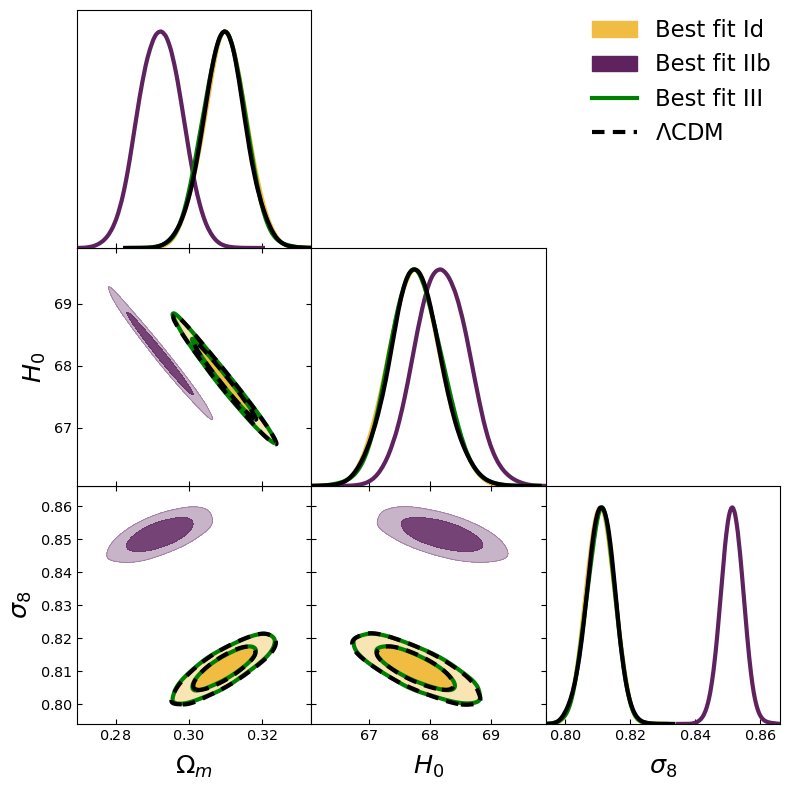}
    \caption{The results of sampling the cosmological parameters while fixing $q$ and $\beta$ to their best fit values found in case Id (yellow), IIb (purple) and case III (green, open). The $\Lambda$CDM result is shown with dashed black lines.}
    \label{fig:bestfits}
\end{figure}

Furthermore, it is possible to produce some additional models which have the same number of degrees of freedom as $\Lambda$CDM, from the best fit values of the parameters found in cases Id, IIb and III. Specifically, we run three additional MCMC chains with $q$ and $\beta$ fixed to their best fit values found (see Table \ref{tab:results}) in those cases. We allow the usual cosmological parameters to vary. The result of these chains is therefore three more Shan--Chen models with the same number of free parameters as $\Lambda$CDM, which we plot in Figure \ref{fig:bestfits}. We can therefore compute the $\Delta \chi^2$ for these additional best fit cases with no need to penalise for the additional degrees of freedom.

The $\Delta \chi^2$ in these best fit cases are: $\Delta \chi^2 = 0.59$ for the best fit case Id; $\Delta \chi^2 = 0.27$ for the best fit case IIb; and $\Delta \chi^2 = 0.61$ for the best fit case III. The best fit Id and III are both very close to $\Lambda$CDM and indeed their original $\Delta\chi^2$ values, representing no improvement in fit. However, the best fit case IIb sees a relatively good improvement in $\chi^2$, and the resulting cosmological parameters are fairly distinct from those of $\Lambda$CDM -- especially $\Omega_m$ and $\sigma_8$ -- as can be seen in Figure \ref{fig:bestfits}. This is because in the best fit case IIb, $q$ is fixed to $-0.1$, whereas in the other two cases $q$ is close to the $\Lambda$CDM limit of zero (recalling that these values of $q$ came from allowing $q$ to vary), thus producing models with cosmological parameters which are more similar to $\Lambda$CDM.

When simply looking at the $\Delta \chi^2$ values for the cases we have studied, a casual reader may conclude that there is very little to choose between the various models. However, this comparison hides a deeper truth about these cases: they describe completely different evolution histories (and futures) for the Universe. To demonstrate this, we can once again plot the behaviour of $w_{\rm int}$, as we did in section \ref{sec:theory}, but this time for the best fit models we have found through our MCMC analysis.

\begin{table}
\centering
\begin{tabular}{Slll} % S prefix allows cells to be stretched by cellspace
	\hline
	\hline
	Case        & $\chi^2$ & $\Delta \chi^2$  \\
	\hline
	Ia ($\Lambda$CDM) & $3831.38$  & $0.0$   \\
	Ib                & $3831.87$  & $0.49$\\
	Ic                & $3831.90$  & $0.52$ \\
	Id                & $3831.98$  & $0.60$  \\
	IIa               & $3833.95$  & $2.57$\\
	IIb               & $3832.76$  & $1.38$  \\
	III               & $3832.20$  & $0.82$ \\
	Best fit Id       & $3831.97$  & $0.59$ \\
	Best fit IIb      & $3831.65$  & $0.27$ \\
	Best fit III      & $3831.99$  & $0.61$ \\
	\hline
\end{tabular}
\caption{The $\chi^2$ and $\Delta \chi^2$ for each case studied.}\label{tab:comparison}
\end{table}

\begin{figure}
    \centering
    \includegraphics[width=\columnwidth]{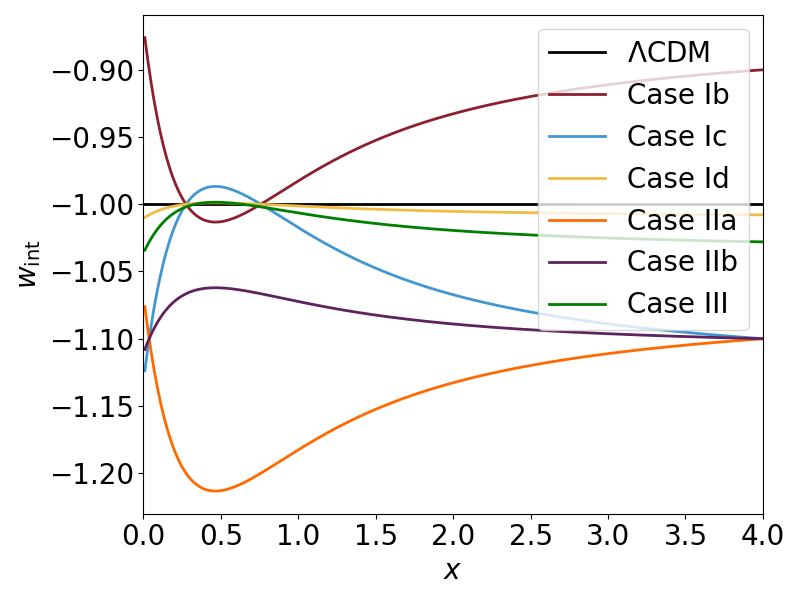}\\
	\includegraphics[width=\columnwidth]{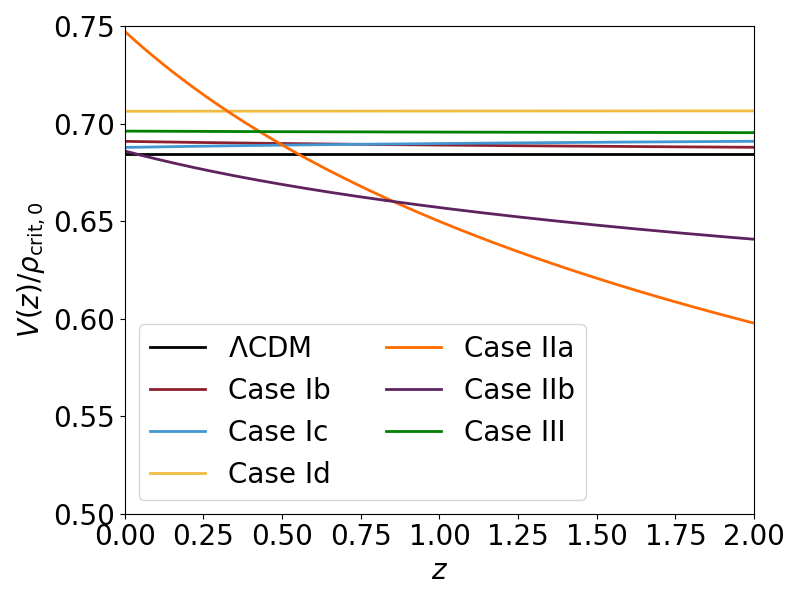} \\
	     \includegraphics[width=\columnwidth]{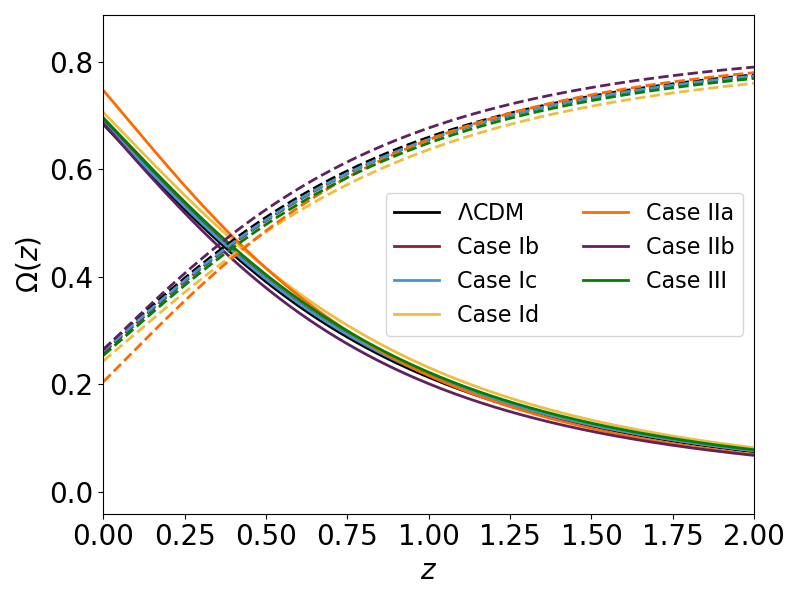}\\
    \caption{Top panel: the behaviour of $w_{\rm int}$ for the best fit models. Middle panel: $V(z)/\rho_{{\rm crit},0}$ for the best fit models. Bottom panel: the evolution of $\Omega_V(z) = 8\pi G V(z)/3H^2$ (solid lines) and $\Omega_m(z)=8\pi G \rho_m(z)/3H^2$ (dashed lines) for the best fit models. Note that at $z=0$, the values of $V(z)/\rho_{{\rm crit},0}$ shown in the middle panel are equivalent to the values of $\Omega_V(z)$ in the bottom panel.}
    \label{fig:v_and_omega}
\end{figure}

\begin{figure*}
    \centering
  
    \includegraphics[width=\columnwidth]{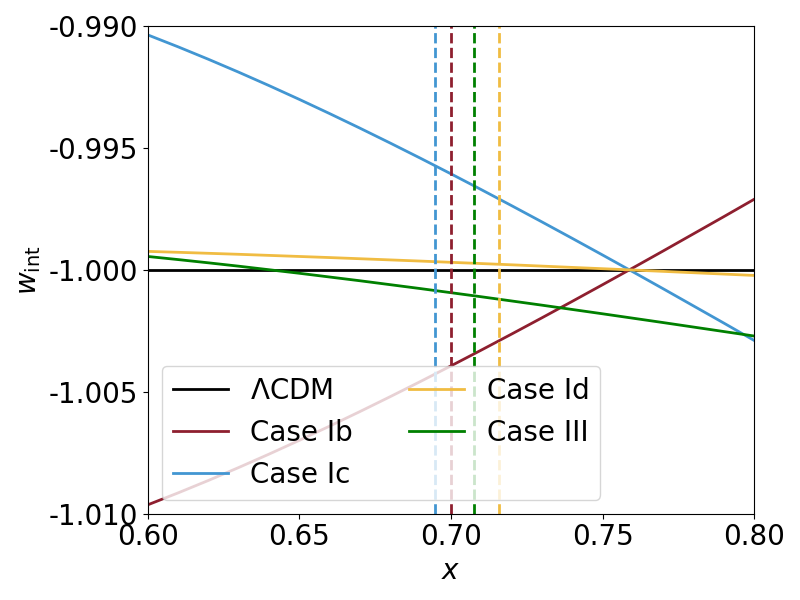}
     \includegraphics[width=\columnwidth]{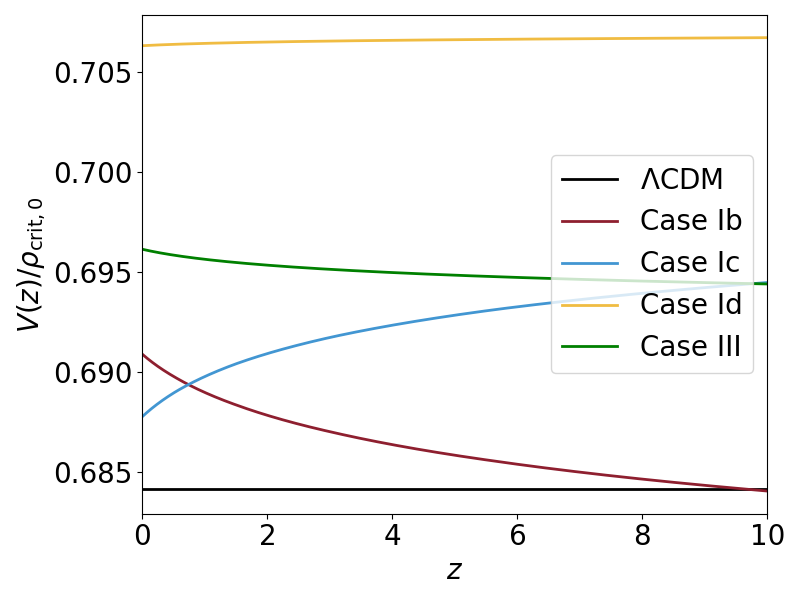}
    \caption{Left panel: $w_{\rm int}$ for case Ib, Ic, Id and III plotted with solid lines, with current values  $V_0=V(0)$   for the same cases, $x_0 = V_0/\bar{\rho}_{\rm crit,0}$,  represented by vertical dashed lines. For cases Ib, Ic and Id, the current value is between the two cosmological constants where $w_{\rm int}=-1$ (the rightmost of the two is shown in the plot); thus in these models the vacuum $V$ evolves between these two constants. For case III the current value is to the right of $w_{\rm int}=-1$ (where the green and the black line cross), hence $V$ in this model is growing away from this constant. Right panel:  $V(z)/\rho_{{\rm crit},0}$ for case Ib, Ic, Id and III. As expected from their current values shown in the left panel, $V$ is growing in cases Ib and III and decaying for cases Ic and Id.}
    \label{fig:v_and_vertical}
\end{figure*}

In the top panel of Figure \ref{fig:v_and_omega} (to be compared with Figure \ref{fig:wint_minmax}),  we show the behaviour of the function $w_{\rm int}(x)$ for all the cases studied in this work, with $q$ and $\beta$ fixed to their best fit values (shown in Table \ref{tab:results}) for those cases in which these parameters were sampled. From this plot, we can see that cases Ic, Id, IIb and III have a maximum in $w_{\rm int}$, while cases Ib and IIa have a minimum. Furthermore, cases IIa and IIb represent models in which $w_{\rm int}<-1$ always, i.e.\ models where the vacuum energy density $V$ is indefinitely growing from zero in the past.
For the cases in which  $w_{\rm int}$ crosses the $-1$ line, i.e. Ib, Ic, Id and III, we can examine the evolution of $V(z)$ in order to determine at which point in $w_{\rm int}$ the particular model sits, i.e. whether $V$ is growing or decaying. 

In the middle panel of Figure \ref{fig:v_and_omega}, we show the evolution of the vacuum energy as a function of redshift, $V(z)$, normalised by the critical density today for each of these models, $\rho_{{\rm crit},0}$, so that the value of $V(z)/\rho_{{\rm crit}, 0}$ at $z=0$ is equivalent to $\Omega_{V}$.  We use \texttt{CAMB} to calculate this evolution, using not only the best fit values of $q$ and $\beta$ in each case, but also the best fit values of the cosmological parameters found in our MCMC analysis. This means that $\rho_{{\rm crit},0}$ is also different for each model.

From this plot, we can see that cases  IIa and IIb both give a very strong growth in $V(z)$.On the other hand, for all of the above mentioned cases where $w_{\rm int}$ crosses the $-1$ line, $V(z)$ is very slowly evolving,  i.e.\  $\rm{d}V/\rm{d}z \approx 0$. Comparing with the bottom panel of Figure \ref{fig:v_and_omega}, we conclude that  for those redshifts where $V(z)$ becomes important, eventually dominating over matter, it is effectively close to a cosmological constant for these four models.

Finally, in the bottom panel of Figure \ref{fig:v_and_omega}, we show the evolution of the matter and vacuum densities in all the cases studied, with $\Omega_V(z)$ for each case shown with a solid line and $\Omega_m(z)$ being shown with dashed lines. For example, from this plot it is clear that case IIa is the most different from $\Lambda$CDM (again shown in black), which is reflected in the uncompetitive $\chi^2$ value for this case.

To enable us to clearly distinguish the four cases which cross $w_{\rm int} = -1$, i.e. cases Ib, Ic, Id and III, we plot them separately in  Figure \ref{fig:v_and_vertical}, where we zoom-in $w_{\rm int}$ in the left panel and we zoom-out $V(z)$ in the right panel, i.e.\  we also increase the redshift range of this plot to emphasise the evolution of $V(z)$ in each case.

In the left panel the vertical lines  represent, for each model, the current value $V_0=V(0)$ of the vacuum $V(z)$, normalised to $\bar{\rho}_{\rm crit, 0}$ (the reference critical energy density in $\Lambda$CDM \citep{Aghanim:2018eyx}),   $x_0 = V_0/ \bar{\rho}_{\rm crit, 0}$. These are the initial values, from $z=0$, in the evolution of $V$, thus their position tells us what the evolution is for each of the four models. We see that for cases Ib, Ic and Id these initial values are between the two cosmological constants for these models, where $w_{\rm int}=-1$ in the top panel of Figure \ref{fig:v_and_omega}, with the rightmost one on the right of the vertical lines in  Figure \ref{fig:v_and_vertical}. Since  for Ic and Id the initial values are where $w_{\rm  int}>-1$, for these models $V$ is currently decreasing from the higher to the lower cosmological constant.  For case Ib, the initial value is where $w_{\rm  int}<-1$, hence $V$ is currently increasing from the lower to the higher cosmological constant. Finally, for case III, the initial value is to the right of the rightmost cosmological constant  in the top panel of Figure \ref{fig:v_and_omega}, which here is the point where the green and the black lines cross; therefore, $V$ is indefinitely growing from this  past asymptotic value. 

In the right panel of in  Figure \ref{fig:v_and_vertical}, we plot $V(z)$  for the four slowly evolving models. The evolution confirms what is is expected from the left panel: case Ib is growing while case Ic is decaying; case Id is also decaying, but at a much slower rate; case III is growing, but again at a slower rate than Ib.

Let us now finally comment on models in which $V$ is always growing, such as cases IIa and IIb: because $w_{\rm int}(x)<-1$ for any $x$, these  should be taken with a pinch of salt, and need to be studied more in detail in order to establish if they are physically viable. Since for these models the coupling \eqref{eq:finalform} is positive and not proportional to the energy density of matter $\rho_c$, it follows from \eqref{mattev} that $\rho_c$ can become negative; in addition, it may well be the case that these models develop some sort of future singularity, see \cite{Ananda:2005xp,Ananda:2006gf,Bamba:2012cp}, and references therein. In this sense, perhaps they cannot be considered good physical models for the entire history of the Universe, but rather a parameterisation of the growth  of the vacuum $V$ (akin to a a phantom-like evolution).

\section{Conclusions} \label{sec:summary}
The primary aim of this paper was to present a novel alternative to the cosmological constant that could be responsible for the accelerating expansion of the Universe, within the general framework of the interacting vacuum scenario. In particular, we were motivated by previous models such as the Chaplygin gas \citep{Kamenshchik:2001cp} or the Van der Waals equation of state \citep{Capozziello:2004ej} to introduce a dark energy which has a physical foundation, and is not purely an ad hoc phenomenological description
such as the Chevallier--Polarski--Linder (CPL) parameterisation of the dark energy equation of state, $w=w_0 + (1-a)w_a$ \citep{Chevallier:2000qy, Linder2003}, in essence a Taylor expansion at low redshift (which is actually less able to distinguish $w_a \neq 0$ than other so-called $w_0$, $w_a$ parameterisations \citep{Colgain:2021pmf}). 

Every dark energy  or unified dark matter energy--momentum tensor can be remapped into vacuum interacting with CDM \citep{Wands:2012vg}.  However, especially in the case of unified dark matter, these models are often subject to severe observational constraints because of the speed of sound of the dark  component \citep{Sandvik:2002jz}, see also \cite{Balbi:2007mz,Pietrobon:2008js,Piattella:2009kt, Piattella:2009da, Gao:2009me, Bertacca:2010mt,DeFelice:2012vd, Wang:2013qy, Li:2018hhy, Li:2019umg}; an advantage of the interacting vacuum formulation is that vacuum does not cluster and, in the specific sub-case of a pure energy exchange we consider here (no momentum exchange in the rest frame of CDM) CDM remains geodesic, as in $\Lambda$CDM.  The generalised Chaplygin gas dark energy model was reconsidered in this fashion by \cite{Bento:2004uh} and \cite{Wands:2012vg}, and \cite{Wang:2013qy} considered observational constraints on the same. 

Following this approach, we cast the fluid dynamical equation of state  originally introduced by \cite{Shan1993} in the context of lattice kinetic theory  into a  CDM--vacuum interaction. This equation of state was considered  as a dark energy model in the restricted case of  a purely FLRW cosmology by \cite{Bini:2014pmk,Bini:2016wqr}. As in previous works \citep{Wands:2012vg, Salvatelli2014, Martinelli:2019dau, Hogg:2020rdp},  we started from  a general covariant formulation of the interacting vacuum scenario, including perturbations, thereby introducing the Shan--Chen interacting vacuum class of  models. 

The Shan--Chen equation of state is nonlinear and characterised by a reference energy scale $\rho_*$, and so is the corresponding interaction, even if the coupling becomes linear \citep{Quercellini:2008vh} at energies well above and below $\rho_*$. Thanks to this nonlinearity a large variety of models can arise, giving  different dynamics, cf.\ \cite{Ananda:2005xp, Ananda:2006gf}: in particular, the vacuum energy density can either grow (from zero or from a past constant value) or decay indefinitely (to zero or to a  constant value), or evolve between two cosmological constants.  We studied a number of models belonging to this class, and placed observational constraints on them. 

Our findings show that the observational data we used (CMB temperature and polarisation, BAO and SNIa  measurements) are compatible with a wide range of models which result in very different cosmologies. Models where $V$ grows indefinitely require further theoretical analysis in order to establish if they lead to a future singularity \citep{Ananda:2005xp, Ananda:2006gf, Bamba:2012cp} but, given that in these models the energy density of CDM becomes negative in the future, more than anything else they can be considered a parameterisation of the vacuum evolution.

The cases studied in this paper are only a small subset of all possible models that exist under the umbrella of Shan--Chen dark energy. We have not given any consideration to the original dark energy fluid as introduced in \cite{Bini:2014pmk}, focusing instead on the Shan--Chen interacting vacuum scenario which we introduced in this work. Even within the Shan--Chen interacting vacuum model, we have limited ourselves to fixing $\rho_* = \bar{\rho}_{\mathrm{crit},0}$, using the  critical energy density of the best fit Planck 2018  $\Lambda$CDM \citep{Aghanim:2018eyx}) as a reference energy scale.
Other possibilities exist, most obviously $\rho_* = \rho_\mathrm{m}(z_{\mathrm{eq}})$. In addition,  the Shan--Chen coupling considered here could be generalised in order to prevent the vacuum $V$ from growing indefinitely and the CDM density $\rho_c$ becoming negative. All of these remain open to exploration in future works.

As we have previously mentioned, besides the desire to find a satisfactory alternative to $\Lambda$, studies of alternative dark energy models such as the Shan--Chen interacting vacuum model presented here can be motivated by their potential to relax the $H_0$ tension. As we found in subsection \ref{subsec:modelcomparison}, cases Ib and Ic have the most competitive $\Delta \chi^2$, apart from the best fit case IIb. We can see from Figure \ref{fig:qfixed} that the values for $H_0$ obtained in Ib and Ic are virtually identical to those in $\Lambda$CDM. As discussed earlier, there also exists a tension between the Planck 2018 value of $\sigma_8$ and that obtained from large scale structure surveys. A resolution of this tension would require a smaller value of $\sigma_8$. The similarity of the cosmological parameters in these particular Shan--Chen interacting vacuum models to those in $\Lambda$CDM means that there is also no relaxation of the $\sigma_8$ tension.

The repeated failures of the interacting vacuum models to cure the $H_0$ tension point to the fact that perhaps this is the wrong line to continue down to try and achieve this particular goal. While it is beyond the scope of this work to explore these ideas, another type of dark energy such as the early dark energy we discussed in \autoref{sec:theory}
could be a more fruitful avenue to explore. Another possibility for future work could be the study of a model which combines both early and late time effects -- \textit{if the sole motivation is a resolution of the $H_0$ tension}. 

Our repurposing of the Shan--Chen equation of state fluid dark energy model as an interacting vacuum was prompted in part by its basis in pre-existing physics, rather than being a purely phenomenological model of a vacuum -- cold dark matter interaction of the type previously considered in the literature. This opens up an interesting philosophical question: should this kind of argument be used more often when constructing alternative dark energy models?

On the one hand, simple models with a strong physical motivation are preferable due to their elegance, and could yield interesting results, as we have seen here. On the other hand, it is important not to become dogmatic when using simplicity as a motivator. For example, a more ``natural'' \citep{Weinberg1989} value of the cosmological constant may be preferable from an aesthetic standpoint -- but such a Universe would be inhospitable to life as we know it \citep{Dijkstra2019}. 

One may ask if the Shan--Chen models presented here are truly physically motivated, or if they too fall under the label of phenomenological. Let us note that while the concept of dark matter is nearly a century old, we still do not have direct evidence for its existence besides the gravitational effects which we attribute to it. Dark energy, either in the form of a cosmological constant or something more exotic like the interacting vacuum considered in this work, is even more puzzling. This leaves us with a huge amount of freedom to construct models to describe these components. In other words, while the
fundamental physics behind dark matter and dark energy remain unknown, the question of whether the Shan--Chen physical model remains valid when transposed from its original context to that of cosmology also remains unanswerable, even more so if the equation of state is turned into an interaction, as we have done here.  Nonetheless,  the Shan-Chen modeling we have introduced is physical at least in the sense that it is covariantly formulated and as such is a description of matter (an interacting vacuum component in our specific case) that  at least in principle could be applied in different contexts, e.g.\ neutron stars, unlike truly phenomenological parameterisations such as CPL  \citep{Chevallier:2000qy, Linder2003} and other similar ad hoc descriptions of evolving dark energy \citep{Colgain:2021pmf}. 

To conclude, the study presented here is a preliminary and non-exhaustive examination of the Shan--Chen interacting vacuum, and many unexplored combinations of parameters remain. In this analysis, we found that no particular Shan--Chen interacting vacuum model of the cases studied here is very competitive with $\Lambda$CDM when performing a model comparison. Overall, our analysis shows it is generally difficult for these particular interacting dark energy models to be both a better fit to the data than $\Lambda$CDM and simultaneously resolve the tensions which persist in cosmology.

\section*{Acknowledgements}
The Python libraries Matplotlib \citep{Hunter:2007} and NumPy \citep{harris2020array} were used for some of the visualisation and analysis presented in this paper. NBH was supported by UK STFC studentship ST/N504245/1 for part of this work, and partly by a postdoctoral position funded through two ``la Caixa'' Foundation fellowships (ID 100010434), with fellowship codes LCF/BQ/PI19/11690015 and LCF/BQ/PI19/11690018. NBH also received financial support from a \href{https://www.gresearch.co.uk/}{G-Research} grant. MB is supported by UK STFC Grant No. ST/S000550/1. Numerical computations were done on the Sciama High Performance Compute (HPC) cluster which is supported by the ICG, SEPNet and the University of Portsmouth. 

\section*{Data Availability}
The data underlying this article will be shared on reasonable request to the corresponding author.

\section*{CRediT authorship contribution statement}
\textbf{Natalie B. Hogg:} Software, formal analysis, investigation, writing -- original draft, visualisation. \textbf{Marco Bruni:} Conceptualisation, formal analysis,  writing -- review and editing, supervision.

\bibliographystyle{mnras}
\bibliography{shanchen} 

\begin{thebibliography}{}
\makeatletter
\relax
\def\mn@urlcharsother{\let\do\@makeother \do\$\do\&\do\#\do\^\do\_\do\%\do\~}
\def\mn@doi{\begingroup\mn@urlcharsother \@ifnextchar [ {\mn@doi@}
  {\mn@doi@[]}}
\def\mn@doi@[#1]#2{\def\@tempa{#1}\ifx\@tempa\@empty \href
  {http://dx.doi.org/#2} {doi:#2}\else \href {http://dx.doi.org/#2} {#1}\fi
  \endgroup}
\def\mn@eprint#1#2{\mn@eprint@#1:#2::\@nil}
\def\mn@eprint@arXiv#1{\href {http://arxiv.org/abs/#1} {{\tt arXiv:#1}}}
\def\mn@eprint@dblp#1{\href {http://dblp.uni-trier.de/rec/bibtex/#1.xml}
  {dblp:#1}}
\def\mn@eprint@#1:#2:#3:#4\@nil{\def\@tempa {#1}\def\@tempb {#2}\def\@tempc
  {#3}\ifx \@tempc \@empty \let \@tempc \@tempb \let \@tempb \@tempa \fi \ifx
  \@tempb \@empty \def\@tempb {arXiv}\fi \@ifundefined
  {mn@eprint@\@tempb}{\@tempb:\@tempc}{\expandafter \expandafter \csname
  mn@eprint@\@tempb\endcsname \expandafter{\@tempc}}}

\bibitem[\protect\citeauthoryear{Abbott et~al.}{Abbott
  et~al.}{2020}]{Abbott:2020knk}
Abbott T.,  et~al., 2020, \mn@doi [Physical Review D]
  {10.1103/PhysRevD.102.023509}, 102, 023509

\bibitem[\protect\citeauthoryear{Aghanim et~al.}{Aghanim
  et~al.}{2020}]{Aghanim:2018eyx}
Aghanim N.,  et~al., 2020, \mn@doi [Astronomy \& Astrophysics]
  {10.1051/0004-6361/201833910}, 641, A6

\bibitem[\protect\citeauthoryear{Alam et~al.}{Alam et~al.}{2017}]{Alam:2016hwk}
Alam S.,  et~al., 2017, \mn@doi [Monthly Notices of the Royal Astronomical
  Society] {10.1093/mnras/stx721}, 470, 2617

\bibitem[\protect\citeauthoryear{Amendola}{Amendola}{2000}]{Amendola1999}
Amendola L.,  2000, \mn@doi [Physical Review D] {10.1103/PhysRevD.62.043511},
  62, 043511

\bibitem[\protect\citeauthoryear{Ananda \& Bruni}{Ananda \&
  Bruni}{2006a}]{Ananda:2005xp}
Ananda K.~N.,  Bruni M.,  2006a, \mn@doi [Phys. Rev. D]
  {10.1103/PhysRevD.74.023523}, 74, 023523

\bibitem[\protect\citeauthoryear{Ananda \& Bruni}{Ananda \&
  Bruni}{2006b}]{Ananda:2006gf}
Ananda K.~N.,  Bruni M.,  2006b, \mn@doi [Phys. Rev. D]
  {10.1103/PhysRevD.74.023524}, 74, 023524

\bibitem[\protect\citeauthoryear{Armendariz-Picon, Mukhanov  \&
  Steinhardt}{Armendariz-Picon et~al.}{2001}]{Amendariz2001}
Armendariz-Picon C.,  Mukhanov V.,   Steinhardt P.~J.,  2001, \mn@doi [Phys.
  Rev. D] {10.1103/PhysRevD.63.103510}, 63, 103510

\bibitem[\protect\citeauthoryear{Balbi, Bruni  \& Quercellini}{Balbi
  et~al.}{2007}]{Balbi:2007mz}
Balbi A.,  Bruni M.,   Quercellini C.,  2007, \mn@doi [Phys. Rev. D]
  {10.1103/PhysRevD.76.103519}, 76, 103519

\bibitem[\protect\citeauthoryear{Bamba, Capozziello, Nojiri  \& Odintsov}{Bamba
  et~al.}{2012}]{Bamba:2012cp}
Bamba K.,  Capozziello S.,  Nojiri S.,   Odintsov S.~D.,  2012, \mn@doi
  [Astrophys. Space Sci.] {10.1007/s10509-012-1181-8}, 342, 155

\bibitem[\protect\citeauthoryear{Banerjee, Cai, Heisenberg, Colg\'ain,
  Sheikh-Jabbari  \& Yang}{Banerjee et~al.}{2021}]{Banerjee:2020xcn}
Banerjee A.,  Cai H.,  Heisenberg L.,  Colg\'ain E.~O.,  Sheikh-Jabbari M.~M.,
   Yang T.,  2021, \mn@doi [Phys. Rev. D] {10.1103/PhysRevD.103.L081305}, 103,
  L081305

\bibitem[\protect\citeauthoryear{Benevento, Hu  \& Raveri}{Benevento
  et~al.}{2020}]{Benevento:2020fev}
Benevento G.,  Hu W.,   Raveri M.,  2020, \mn@doi [Phys. Rev. D]
  {10.1103/PhysRevD.101.103517}, 101, 103517

\bibitem[\protect\citeauthoryear{Bento, Bertolami  \& Sen}{Bento
  et~al.}{2004}]{Bento:2004uh}
Bento M.~C.,  Bertolami O.,   Sen A.~A.,  2004, \mn@doi [Phys. Rev. D]
  {10.1103/PhysRevD.70.083519}, 70, 083519

\bibitem[\protect\citeauthoryear{Bertacca, Bruni, Piattella  \&
  Pietrobon}{Bertacca et~al.}{2011}]{Bertacca:2010mt}
Bertacca D.,  Bruni M.,  Piattella O.~F.,   Pietrobon D.,  2011, \mn@doi
  [Journal of Cosmology and Astroparticle Physics]
  {10.1088/1475-7516/2011/02/018}, 1102, 018

\bibitem[\protect\citeauthoryear{{Beutler} et~al.,}{{Beutler}
  et~al.}{2011}]{Beutler2011}
{Beutler} F.,  et~al., 2011, \mn@doi [Monthly Notices of the Royal Astronomical
  Society] {10.1111/j.1365-2966.2011.19250.x}, \href
  {http://adsabs.harvard.edu/abs/2011MNRAS.416.3017B} {416, 3017}

\bibitem[\protect\citeauthoryear{Bini, Geralico, Gregoris  \& Succi}{Bini
  et~al.}{2013}]{Bini:2014pmk}
Bini D.,  Geralico A.,  Gregoris D.,   Succi S.,  2013, \mn@doi [Phys. Rev. D]
  {10.1103/PhysRevD.88.063007}, 88, 063007

\bibitem[\protect\citeauthoryear{Bini, Esposito  \& Geralico}{Bini
  et~al.}{2016}]{Bini:2016wqr}
Bini D.,  Esposito G.,   Geralico A.,  2016, \mn@doi [Phys. Rev. D]
  {10.1103/PhysRevD.93.023511}, 93, 023511

\bibitem[\protect\citeauthoryear{Camarena \& Marra}{Camarena \&
  Marra}{2021}]{Camarena:2021jlr}
Camarena D.,  Marra V.,  2021, \mn@doi [Mon. Not. Roy. Astron. Soc.]
  {10.1093/mnras/stab1200}, 504, 5164

\bibitem[\protect\citeauthoryear{Capozziello, Cardone, Carloni, De~Martino,
  Falanga, Troisi  \& Bruni}{Capozziello et~al.}{2005}]{Capozziello:2004ej}
Capozziello S.,  Cardone V.~F.,  Carloni S.,  De~Martino S.,  Falanga M.,
  Troisi A.,   Bruni M.,  2005, \mn@doi [JCAP] {10.1088/1475-7516/2005/04/005},
  04, 005

\bibitem[\protect\citeauthoryear{Chevallier \& Polarski}{Chevallier \&
  Polarski}{2001}]{Chevallier:2000qy}
Chevallier M.,  Polarski D.,  2001, \mn@doi [Int. J. Mod. Phys. D]
  {10.1142/S0218271801000822}, 10, 213

\bibitem[\protect\citeauthoryear{Colg\'ain, Sheikh-Jabbari  \& Yin}{Colg\'ain
  et~al.}{2021}]{Colgain:2021pmf}
Colg\'ain E.~O.,  Sheikh-Jabbari M.~M.,   Yin L.,  2021, \mn@doi [Phys. Rev. D]
  {10.1103/PhysRevD.104.023510}, 104, 023510

\bibitem[\protect\citeauthoryear{{Copeland}, {Sami}  \& {Tsujikawa}}{{Copeland}
  et~al.}{2006}]{Copeland2006}
{Copeland} E.~J.,  {Sami} M.,   {Tsujikawa} S.,  2006, \mn@doi [International
  Journal of Modern Physics D] {10.1142/S021827180600942X}, \href
  {https://ui.adsabs.harvard.edu/abs/2006IJMPD..15.1753C} {15, 1753}

\bibitem[\protect\citeauthoryear{De~Felice, Nesseris  \& Tsujikawa}{De~Felice
  et~al.}{2012}]{DeFelice:2012vd}
De~Felice A.,  Nesseris S.,   Tsujikawa S.,  2012, \mn@doi [Journal of
  Cosmology and Astroparticle Physics] {10.1088/1475-7516/2012/05/029}, 05, 029

\bibitem[\protect\citeauthoryear{Di~Valentino et~al.,}{Di~Valentino
  et~al.}{2021a}]{DiValentino:2021izs}
Di~Valentino E.,  et~al., 2021a, \mn@doi [Class. Quant. Grav.]
  {10.1088/1361-6382/ac086d}, 38, 153001

\bibitem[\protect\citeauthoryear{Di~Valentino et~al.}{Di~Valentino
  et~al.}{2021b}]{DiValentino:2020zio}
Di~Valentino E.,  et~al., 2021b, \mn@doi [Astropart. Phys.]
  {10.1016/j.astropartphys.2021.102605}, 131, 102605

\bibitem[\protect\citeauthoryear{Dijkstra}{Dijkstra}{2019}]{Dijkstra2019}
Dijkstra C.~D.,  2019, Master's thesis, University of Groningen, \url
  {arxiv.org/abs/1906.03036}

\bibitem[\protect\citeauthoryear{Dodge}{Dodge}{2008}]{chi2book}
Dodge Y.,  2008, Chi-Square Table.
Springer New York, New York, NY, pp 76--77,
  \mn@doi{10.1007/978-0-387-32833-1_56}, \url
  {https://doi.org/10.1007/978-0-387-32833-1_56}

\bibitem[\protect\citeauthoryear{Efstathiou}{Efstathiou}{2021}]{Efstathiou:2021ocp}
Efstathiou G.,  2021, \mn@doi [Mon. Not. Roy. Astron. Soc.]
  {10.1093/mnras/stab1588}, 505, 3866

\bibitem[\protect\citeauthoryear{Eke, Navarro  \& Frenk}{Eke
  et~al.}{1998}]{Eke1998}
Eke V.~R.,  Navarro J.~F.,   Frenk C.~S.,  1998, \mn@doi [The Astrophysical
  Journal] {10.1086/306008}, 503, 569

\bibitem[\protect\citeauthoryear{Freedman et~al.}{Freedman
  et~al.}{2019}]{Freedman:2019jwv}
Freedman W.~L.,  et~al., 2019, \mn@doi [The Astrophysical Journal]
  {10.3847/1538-4357/ab2f73}, 882, 34

\bibitem[\protect\citeauthoryear{Freese \& Winkler}{Freese \&
  Winkler}{2021}]{Freese:2021rjq}
Freese K.,  Winkler M.~W.,  2021, {Chain Early Dark Energy: Solving the Hubble
  Tension and Explaining Today's Dark Energy} (\mn@eprint {arXiv} {2102.13655})

\bibitem[\protect\citeauthoryear{Frenk \& White}{Frenk \&
  White}{2012}]{Frenk:2012ph}
Frenk C.,  White S.~D.,  2012, \mn@doi [Annalen der Physik]
  {10.1002/andp.201200212}, 524, 507

\bibitem[\protect\citeauthoryear{Gao, Kunz, Liddle  \& Parkinson}{Gao
  et~al.}{2010}]{Gao:2009me}
Gao C.,  Kunz M.,  Liddle A.~R.,   Parkinson D.,  2010, \mn@doi [Physical
  Review D] {10.1103/PhysRevD.81.043520}, 81, 043520

\bibitem[\protect\citeauthoryear{Haridasu, Viel  \& Vittorio}{Haridasu
  et~al.}{2021}]{Haridasu:2020pms}
Haridasu B.~S.,  Viel M.,   Vittorio N.,  2021, \mn@doi [Phys. Rev. D]
  {10.1103/PhysRevD.103.063539}, 103, 063539

\bibitem[\protect\citeauthoryear{Harris et~al.,}{Harris
  et~al.}{2020}]{harris2020array}
Harris C.~R.,  et~al., 2020, \mn@doi [Nature] {10.1038/s41586-020-2649-2}, 585,
  357

\bibitem[\protect\citeauthoryear{Heymans et~al.}{Heymans
  et~al.}{2021}]{Heymans2020}
Heymans C.,  et~al., 2021, \mn@doi [Astron. Astrophys.]
  {10.1051/0004-6361/202039063}, 646, A140

\bibitem[\protect\citeauthoryear{Hill, McDonough, Toomey  \& Alexander}{Hill
  et~al.}{2020}]{Hill:2020osr}
Hill J.~C.,  McDonough E.,  Toomey M.~W.,   Alexander S.,  2020, \mn@doi [Phys.
  Rev. D] {10.1103/PhysRevD.102.043507}, 102, 043507

\bibitem[\protect\citeauthoryear{Hogg, Bruni, Crittenden, Martinelli  \&
  Peirone}{Hogg et~al.}{2020}]{Hogg:2020rdp}
Hogg N.~B.,  Bruni M.,  Crittenden R.,  Martinelli M.,   Peirone S.,  2020,
  \mn@doi [Phys. Dark Univ.] {10.1016/j.dark.2020.100583}, 29, 100583

\bibitem[\protect\citeauthoryear{Howlett, Lewis, Hall  \& Challinor}{Howlett
  et~al.}{2012}]{Howlett:2012mh}
Howlett C.,  Lewis A.,  Hall A.,   Challinor A.,  2012, \mn@doi [Journal of
  Cosmology and Astroparticle Physics] {10.1088/1475-7516/2012/04/027}, 1204,
  027

\bibitem[\protect\citeauthoryear{Hunter}{Hunter}{2007}]{Hunter:2007}
Hunter J.~D.,  2007, \mn@doi [Computing in Science \& Engineering]
  {10.1109/MCSE.2007.55}, 9, 90

\bibitem[\protect\citeauthoryear{Ivanov, McDonough, Hill, Simonovi\'c, Toomey,
  Alexander  \& Zaldarriaga}{Ivanov et~al.}{2020}]{Ivanov:2020ril}
Ivanov M.~M.,  McDonough E.,  Hill J.~C.,  Simonovi\'c M.,  Toomey M.~W.,
  Alexander S.,   Zaldarriaga M.,  2020, \mn@doi [Phys. Rev. D]
  {10.1103/PhysRevD.102.103502}, 102, 103502

\bibitem[\protect\citeauthoryear{Kamenshchik, Moschella  \&
  Pasquier}{Kamenshchik et~al.}{2001}]{Kamenshchik:2001cp}
Kamenshchik A.~Y.,  Moschella U.,   Pasquier V.,  2001, \mn@doi [Physics
  Letters B] {10.1016/S0370-2693(01)00571-8}, 511, 265

\bibitem[\protect\citeauthoryear{Karwal, Raveri, Jain, Khoury  \&
  Trodden}{Karwal et~al.}{2021}]{Karwal:2021vpk}
Karwal T.,  Raveri M.,  Jain B.,  Khoury J.,   Trodden M.,  2021, {Chameleon
  Early Dark Energy and the Hubble Tension} (\mn@eprint {arXiv} {2106.13290})

\bibitem[\protect\citeauthoryear{Kunz, Nesseris  \& Sawicki}{Kunz
  et~al.}{2016}]{Kunz:2016yqy}
Kunz M.,  Nesseris S.,   Sawicki I.,  2016, \mn@doi [Physical Review D]
  {10.1103/PhysRevD.94.023510}, 94, 023510

\bibitem[\protect\citeauthoryear{Lewis}{Lewis}{2013}]{Lewis:2013hha}
Lewis A.,  2013, \mn@doi [Physical Review D] {10.1103/PhysRevD.87.103529}, 87,
  103529

\bibitem[\protect\citeauthoryear{Lewis}{Lewis}{2019}]{Lewis:2019xzd}
Lewis A.,  2019, {GetDist: a Python package for analysing Monte Carlo samples}
  (\mn@eprint {arXiv} {1910.13970})

\bibitem[\protect\citeauthoryear{Lewis \& Bridle}{Lewis \&
  Bridle}{2002}]{Lewis:2002ah}
Lewis A.,  Bridle S.,  2002, \mn@doi [Physical Review D]
  {10.1103/PhysRevD.66.103511}, 66, 103511

\bibitem[\protect\citeauthoryear{Lewis, Challinor  \& Lasenby}{Lewis
  et~al.}{2000}]{Lewis:1999bs}
Lewis A.,  Challinor A.,   Lasenby A.,  2000, \mn@doi [The Astrophysical
  Journal] {10.1086/309179}, 538, 473

\bibitem[\protect\citeauthoryear{Li, Yang  \& Wu}{Li et~al.}{2018}]{Li:2018hhy}
Li H.,  Yang W.,   Wu Y.,  2018, \mn@doi [Physics of the Dark Universe]
  {10.1016/j.dark.2018.09.001}, 22, 60

\bibitem[\protect\citeauthoryear{Li, Yang  \& Gai}{Li
  et~al.}{2019}]{Li:2019umg}
Li H.,  Yang W.,   Gai L.,  2019, \mn@doi [Astronomy \& Astrophysics]
  {10.1051/0004-6361/201833836}, 623, A28

\bibitem[\protect\citeauthoryear{Linder}{Linder}{2003}]{Linder2003}
Linder E.~V.,  2003, \mn@doi [Phys. Rev. Lett.]
  {10.1103/PhysRevLett.90.091301}, 90, 091301

\bibitem[\protect\citeauthoryear{Linder}{Linder}{2007}]{Linder2007}
Linder E.~V.,  2007, General Relativity and Gravitation, 40, 329

\bibitem[\protect\citeauthoryear{{Martin}}{{Martin}}{2012}]{Martin2012}
{Martin} J.,  2012, \mn@doi [Comptes Rendus Physique]
  {10.1016/j.crhy.2012.04.008}, \href
  {https://ui.adsabs.harvard.edu/abs/2012CRPhy..13..566M} {13, 566}

\bibitem[\protect\citeauthoryear{{Martinelli}, {Hogg}, {Peirone}, {Bruni}  \&
  {Wands}}{{Martinelli} et~al.}{2019}]{Martinelli:2019dau}
{Martinelli} M.,  {Hogg} N.~B.,  {Peirone} S.,  {Bruni} M.,   {Wands} D.,
  2019, \mn@doi [Monthly Notices of the Royal Astronomical Society]
  {10.1093/mnras/stz1915}, \href
  {https://ui.adsabs.harvard.edu/abs/2019MNRAS.488.3423M} {488, 3423}

\bibitem[\protect\citeauthoryear{Niedermann \& Sloth}{Niedermann \&
  Sloth}{2020}]{Niedermann:2020dwg}
Niedermann F.,  Sloth M.~S.,  2020, \mn@doi [Phys. Rev. D]
  {10.1103/PhysRevD.102.063527}, 102, 063527

\bibitem[\protect\citeauthoryear{Park, Raveri  \& Jain}{Park
  et~al.}{2021}]{Park:2021jmi}
Park M.,  Raveri M.,   Jain B.,  2021, \mn@doi [Phys. Rev. D]
  {10.1103/PhysRevD.103.103530}, 103, 103530

\bibitem[\protect\citeauthoryear{{Perlmutter} et~al.}{{Perlmutter}
  et~al.}{1999}]{Perlmutter1999}
{Perlmutter} S.,  et~al., 1999, \mn@doi [The Astrophysical Journal]
  {10.1086/307221}, \href {http://adsabs.harvard.edu/abs/1999ApJ...517..565P}
  {517, 565}

\bibitem[\protect\citeauthoryear{Pesce et~al.}{Pesce
  et~al.}{2020}]{Pesce:2020xfe}
Pesce D.,  et~al., 2020, \mn@doi [The Astrophysical Journal Letters]
  {10.3847/2041-8213/ab75f0}, 891, L1

\bibitem[\protect\citeauthoryear{Piattella}{Piattella}{2010}]{Piattella:2009da}
Piattella O.~F.,  2010, \mn@doi [Journal of Cosmology and Astroparticle
  Physics] {10.1088/1475-7516/2010/03/012}, 03, 012

\bibitem[\protect\citeauthoryear{Piattella, Bertacca, Bruni  \&
  Pietrobon}{Piattella et~al.}{2010}]{Piattella:2009kt}
Piattella O.~F.,  Bertacca D.,  Bruni M.,   Pietrobon D.,  2010, \mn@doi
  [Journal of Cosmology and Astroparticle Physics]
  {10.1088/1475-7516/2010/01/014}, 1001, 014

\bibitem[\protect\citeauthoryear{Pietrobon, Balbi, Bruni  \&
  Quercellini}{Pietrobon et~al.}{2008}]{Pietrobon:2008js}
Pietrobon D.,  Balbi A.,  Bruni M.,   Quercellini C.,  2008, \mn@doi [Phys.
  Rev. D] {10.1103/PhysRevD.78.083510}, 78, 083510

\bibitem[\protect\citeauthoryear{Poulin, Smith, Karwal  \& Kamionkowski}{Poulin
  et~al.}{2019}]{Poulin:2018cxd}
Poulin V.,  Smith T.~L.,  Karwal T.,   Kamionkowski M.,  2019, \mn@doi
  [Physical Review Letters] {10.1103/PhysRevLett.122.221301}, 122, 221301

\bibitem[\protect\citeauthoryear{Quercellini, Bruni, Balbi  \&
  Pietrobon}{Quercellini et~al.}{2008}]{Quercellini:2008vh}
Quercellini C.,  Bruni M.,  Balbi A.,   Pietrobon D.,  2008, \mn@doi [Phys.
  Rev. D] {10.1103/PhysRevD.78.063527}, 78, 063527

\bibitem[\protect\citeauthoryear{{Riess} et~al.}{{Riess}
  et~al.}{1998}]{Riess1998}
{Riess} A.~G.,  et~al., 1998, \mn@doi [The Astronomical Journal]
  {10.1086/300499}, \href {http://adsabs.harvard.edu/abs/1998AJ....116.1009R}
  {116, 1009}

\bibitem[\protect\citeauthoryear{{Riess}, {Casertano}, {Yuan}, {Macri}  \&
  {Scolnic}}{{Riess} et~al.}{2019}]{Riess:2019cxk}
{Riess} A.~G.,  {Casertano} S.,  {Yuan} W.,  {Macri} L.~M.,   {Scolnic} D.,
  2019, \mn@doi [The Astrophysical Journal] {10.3847/1538-4357/ab1422}, \href
  {https://ui.adsabs.harvard.edu/abs/2019ApJ...876...85R} {876, 85}

\bibitem[\protect\citeauthoryear{Ross, Samushia, Howlett, Percival, Burden  \&
  Manera}{Ross et~al.}{2015}]{Ross:2014qpa}
Ross A.~J.,  Samushia L.,  Howlett C.,  Percival W.~J.,  Burden A.,   Manera
  M.,  2015, \mn@doi [Mon. Not. Roy. Astron. Soc.] {10.1093/mnras/stv154}, 449,
  835

\bibitem[\protect\citeauthoryear{{Sakai}}{{Sakai}}{1999}]{Sakai1999}
{Sakai} S.,  1999, {The Tip of the Red Giant Branch as a Population II Distance
  Indicator}.
p.~48

\bibitem[\protect\citeauthoryear{Salvatelli, Said, Bruni, Melchiorri  \&
  Wands}{Salvatelli et~al.}{2014}]{Salvatelli2014}
Salvatelli V.,  Said N.,  Bruni M.,  Melchiorri A.,   Wands D.,  2014, \mn@doi
  [Physical Review Letters] {10.1103/PhysRevLett.113.181301}, 113, 181301

\bibitem[\protect\citeauthoryear{Sandvik, Tegmark, Zaldarriaga  \&
  Waga}{Sandvik et~al.}{2004}]{Sandvik:2002jz}
Sandvik H.,  Tegmark M.,  Zaldarriaga M.,   Waga I.,  2004, \mn@doi [Physical
  Review D] {10.1103/PhysRevD.69.123524}, 69, 123524

\bibitem[\protect\citeauthoryear{Scolnic et~al.}{Scolnic
  et~al.}{2018}]{Scolnic:2017caz}
Scolnic D.~M.,  et~al., 2018, \mn@doi [Astrophys. J.]
  {10.3847/1538-4357/aab9bb}, 859, 101

\bibitem[\protect\citeauthoryear{Shan \& Chen}{Shan \& Chen}{1993}]{Shan1993}
Shan X.,  Chen H.,  1993, \mn@doi [Phys. Rev. E] {10.1103/PhysRevE.47.1815},
  47, 1815

\bibitem[\protect\citeauthoryear{Sofue \& Rubin}{Sofue \&
  Rubin}{2001}]{Sofue:2000jx}
Sofue Y.,  Rubin V.,  2001, \mn@doi [Annual Review of Astronomy and
  Astrophysics] {10.1146/annurev.astro.39.1.137}, 39, 137

\bibitem[\protect\citeauthoryear{Wands, De-Santiago  \& Wang}{Wands
  et~al.}{2012}]{Wands:2012vg}
Wands D.,  De-Santiago J.,   Wang Y.,  2012, \mn@doi [Classical and Quantum
  Gravity] {10.1088/0264-9381/29/14/145017}, 29, 145017

\bibitem[\protect\citeauthoryear{Wang, Wands, Xu, De-Santiago  \& Hojjati}{Wang
  et~al.}{2013}]{Wang:2013qy}
Wang Y.,  Wands D.,  Xu L.,  De-Santiago J.,   Hojjati A.,  2013, \mn@doi
  [Physical Review D] {10.1103/PhysRevD.87.083503}, 87, 083503

\bibitem[\protect\citeauthoryear{Weinberg}{Weinberg}{1989}]{Weinberg1989}
Weinberg S.,  1989, Reviews of Modern Physics, 61, 1

\bibitem[\protect\citeauthoryear{Wong et~al.}{Wong et~al.}{2020}]{Wong:2019kwg}
Wong K.~C.,  et~al., 2020, \mn@doi [Monthly Notices of the Royal Astronomical
  Society] {10.1093/mnras/stz3094}, 498

\bibitem[\protect\citeauthoryear{{Zwicky}}{{Zwicky}}{1933}]{Zwicky1933}
{Zwicky} F.,  1933, Helvetica Physica Acta, \href
  {https://ui.adsabs.harvard.edu/abs/1933AcHPh...6..110Z} {6, 110}

\bibitem[\protect\citeauthoryear{de Putter \& Linder}{de~Putter \&
  Linder}{2007}]{Putter2007}
de Putter R.,  Linder E.~V.,  2007, Astroparticle Physics, 28, 263

\makeatother
\end{thebibliography}

\appendix
\section{Additional results} \label{sec:appendix}
In this appendix, we present two additional sets of results. We emphasise that these are presented for the sake of completeness and are in line with the rest of our findings and conclusions about the Shan--Chen interacting vacuum.

\begin{figure}
    \centering
    \includegraphics[width=\columnwidth]{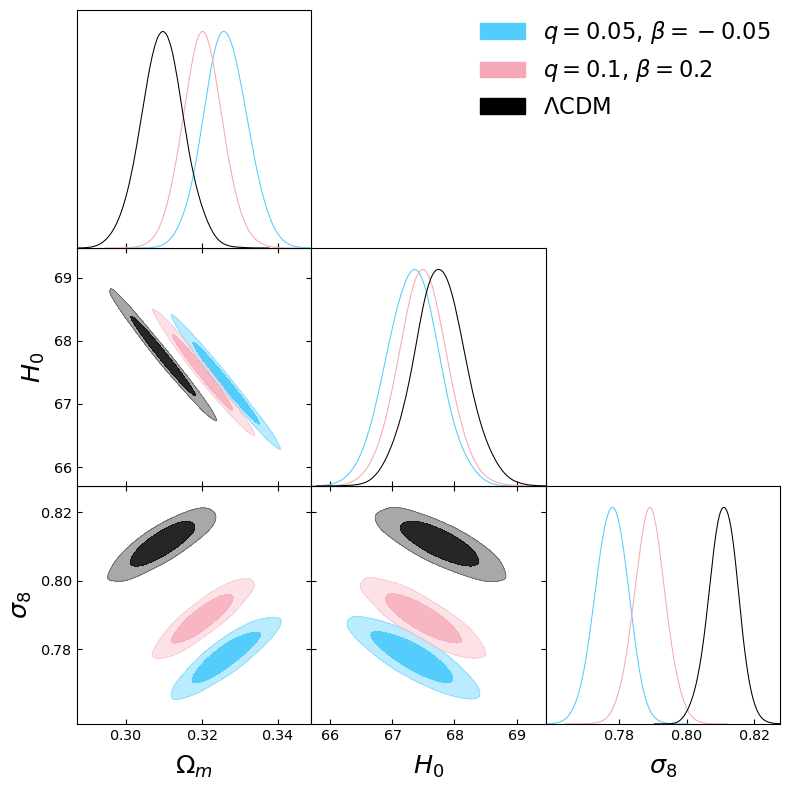}
    \caption{Results of sampling the cosmological parameters while keeping $q=0.05$ and $\beta=-0.05$ (light blue), and when $q=0.1$ and $\beta=0.2$ (pink). The $\Lambda$CDM result is shown in black.}
    \label{fig:additional_results}
\end{figure}

In Figure \ref{fig:additional_results}, we plot the results from sampling the cosmological parameters when $q=0.05$ and $\beta=-0.05$ (light blue), and when $q=0.1$ and $\beta=0.2$ (pink). We can see that in both of these cases, the error bars on the parameters (inferred from the 1 and 2 $\sigma$ regions of the contours) are equivalent to those in $\Lambda$CDM (black), which is the same as what we find in the main cases studied in this work when we keep both $q$ and $\beta$ fixed (i.e. cases Ib, Ic and IIa). We can also see that in both of these cases, the $H_0$ tension is largely unaffected, while the $\sigma_8$ tension is somewhat relaxed. The cases shown here have a $\Delta \chi^2$ of 1.49 and  0.97 respectively, showing that they are both worse fits to the data than $\Lambda$CDM.

% Don't change these lines
\bsp	% typesetting comment
\label{lastpage}
\end{document}